\documentclass{article}
\usepackage[utf8]{inputenc}
\usepackage{overpic}

\usepackage{amsmath}
\usepackage{amssymb}
\usepackage{graphicx}
\usepackage{float}
\usepackage{caption}
\usepackage{subcaption}
\usepackage{multirow}
\usepackage{comment}
\usepackage{url}
\usepackage{hyperref}
\usepackage{dashrule}
\usepackage[dvipsnames]{xcolor}
\usepackage{siunitx}
\usepackage{pict2e}
\usepackage[colorinlistoftodos]{todonotes}
\newcommand{\ian}[1]{\todo[inline,color=purple!40]{Ian: #1}}

\newcommand{\zechariah}[1]{\todo[inline,color=red!40]{Zechariah: #1}}

\newcommand{\con}{\mathcal{C}}
\title{Predicting the Spatially Varying Infection Risk in Indoor Spaces Using an Efficient Airborne Transmission Model}
\author{Zechariah Lau$^{1,2}$, Ian M.~Griffiths$^2$,\\ Aaron English$^{1,3}$, Katerina Kaouri$^1\footnote{kaourik@cardiff.ac.uk}$}
\date{\small{$^1$ School of Mathematics, Cardiff University, CF24 4AG, \\ $^2$ Mathematical Institute, University of Oxford, OX1 6GG \\
$^3$ Department of Mechanical, Aerospace and Civil Engineering, \\ University of Manchester M13 9PL}}

\begin{document}

\maketitle

\vspace{-7mm}


\begin{abstract}


   We develop a spatially dependent generalisation to the Wells--Riley model~\cite{Riley1978} and its extensions applied to COVID-19, for example \cite{Buonanno2020}, that determines the infection risk due to airborne transmission of viruses. We assume that the concentration of infectious particles is governed by an advection--diffusion--reaction equation with the particles advected by airflow, diffused due to turbulence, emitted by infected people and removed due to the room ventilation, inactivation of the virus and gravitational settling. We consider one asymptomatic or presymptomatic infectious person who breathes or talks, with or without a mask and model a quasi-3D setup that incorporates a recirculating air-conditioning flow. A semi-analytic solution is available and this enables fast simulations. We quantify the effect of ventilation and particle emission rate on the particle concentration, infection risk and the `time to probable infection’ (TTPI). Good agreement with CFD models is achieved. Furthermore, we derive power laws that quantify the effect of ventilation, emission rate and infectiousness of the virus on the TTPI. The model can be easily updated to take into account modified parameter values. This work paves the way for establishing `safe occupancy times' at any location and has direct applicability in mitigating the spread of the COVID-19 pandemic. 
   \end{abstract}

\section{Introduction}
The COVID-19 pandemic has spread rapidly across the globe, with more than 152 million confirmed cases worldwide and over three million deaths~\cite{web:WHO-Dashboard}. The virus causing the pandemic, SARS-CoV-2, is transmitted through virus-carrying respiratory particles, which are released when an infected person coughs, sneezes, talks or breathes \cite{WHO-modes,Jin2020}. Evidence has accumulated that airborne aerosols formed from smaller respiratory droplets evaporating can also transmit the disease \cite{Asadi2020}. In July 2020, 239 scientists signed an open letter appealing for the recognition of airborne transmission \cite{Morawska2020} and in October 2020 the US Centre for Disease Prevention and Control acknowledged airborne transmission and updated their guidelines \cite{CDC-guideline}. In March 2021, Public Health England recommended improving ventilation of indoor spaces as an intervention against COVID-19 \cite{england-vent}.

Models studying the risk of airborne transmission generally fall into one of two types:  Computational Fluid Dynamics (CFD) models and the Wells--Riley model and its extensions. CFD models are useful in studying airborne transmission as they can take into account the details of the room size, geometry, the complex turbulent airflow and the size distribution of the aerosols. Various CFD models been applied to the COVID-19 pandemic to investigated the transport of aerosols 
\cite{Shafaghi2020,Vuorinen2020,Feng2020, Li2020,Birnir2020,Shao2020,Ho2021}. Some of these studies focus on the transport in the short term (less than 5 minutes) \cite{Shafaghi2020,Vuorinen2020,Feng2020}, while others show the build-up of aerosols indoors over an hour \cite{Birnir2020,Shao2020}. The risk of infection can then be estimated from these results \cite{Shao2020,Ho2021}. However, CFD models require specialised software and are computationally demanding.They take a long time to run even for small-sized locations, so they cannot easily be applied to a new location and are hence not suited for `dynamic simulations', which are essential in the current fast-evolving pandemic situation.

On the other hand, the Wells--Riley model and its generalisations such as the Gammaitoni-Nucci extension are built on the Well-Mixed-Room (WMR) assumption, which assumes that the virus-carrying aerosols are instantaneously evenly distributed throughout the room \cite{Riley1978,Wells1955,gammaitoni1997using}. This assumption implies that everyone in the room is equally likely to be infected, regardless of their position. This is a major simplification of the problem as it neglects the complex effects of the air flow on the airborne particles, and the effects of the room geometry. However, the latter models are easy to implement and very fast to run \cite{gammaitoni1997using,web:Jimenez}. The Wells--Riley model and its extensions were quickly applied to make predictions on COVID-19 transmission~\cite{Buonanno2020, web:Jimenez, Dai2020, Sun2020,Lelieveld2020,miller2021transmission,burridge2021}.

In this paper we develop an extension to the Wells--Riley model~\cite{Riley1978,gammaitoni1997using} and determine the \emph{spatially dependent} airborne transmission risk in a room. The model describes the concentration of the particles via an advection--diffusion--reaction equation, and allows fast simulations while still taking into account the effect of the turbulent airflow. Even though our model is less detailed than CFD models, it is able to reproduce the results of more complex CFD simulations, while running quickly on an average PC, without the need of a supercomputer. Hence our model is slightly more computationally demanding than the Wells--Riley model.
The computational simplicity of our model is an asset during the fast decision-making that required in this pandemic as it enables easy application to different locations such as classrooms, healthcare clinics and restaurants.

In Section~\ref{Section:Methodology}, we present our model for the concentration of the airborne particles indoors, produced by a single presymptomatic or asymptomatic spreader of COVID-19 who is breathing or talking in a room with only mechanical ventilation (no open doors or windows). We model the recirculation of the air due to air-conditioning by modelling the concentration at all positions on a closed loop around the room, which provides a quasi-3D setup. In this case, we show that a semi-analytic solution exists for the concentration as a function of space and time. We then use the concentration results in an exponential probability density function that determine the probability of infection. In Section~\ref{Section:Parameters}, we discuss the parameter values in our model and how these may be easily changed once updated knowledge becomes available. 

As many countries are still debating the best way to operate educational spaces in order to minimise infection rates during the ongoing third wave of the pandemic, in Section~\ref{Section:Results and Discussion} we implement the model for an average-sized classroom. We determine the particle concentration and the infection risk in the classroom for four different ventilation scenarios: very poor ventilation, poor ventilation, a pre-pandemic ASHRAE recommended ventilation and a pandemic-updated ASHRAE recommended ventilation. 
We also run the model for the superspreader outbreak at a restaurant in Guangzhou, China that occurred on January 23, 2020 and infected 7--9 people \cite{Lu2020} and find satisfactory agreement with two CFD models that simulated this outbreak.
Furthermore, probing our mathematical model we uncover power-law relationships in locations and times when the walls of the room do not play a role in the viral concentration. Such power laws allow for simple predictions that may be easily implemented in a range of locations. 

We summarise our work and indicate how policy-makers and other decision-makers could use the model in future planning to tackle the pandemic in Section~\ref{Section:Summary and Conclusions}.
\section{Modelling Framework}
\label{Section:Methodology}

\subsection{The Advection--Diffusion--Reaction (ADR) Equation}

\begin{figure}[t]
    \hspace{-10mm}\begin{overpic}[width=1.2\textwidth,tics=10]{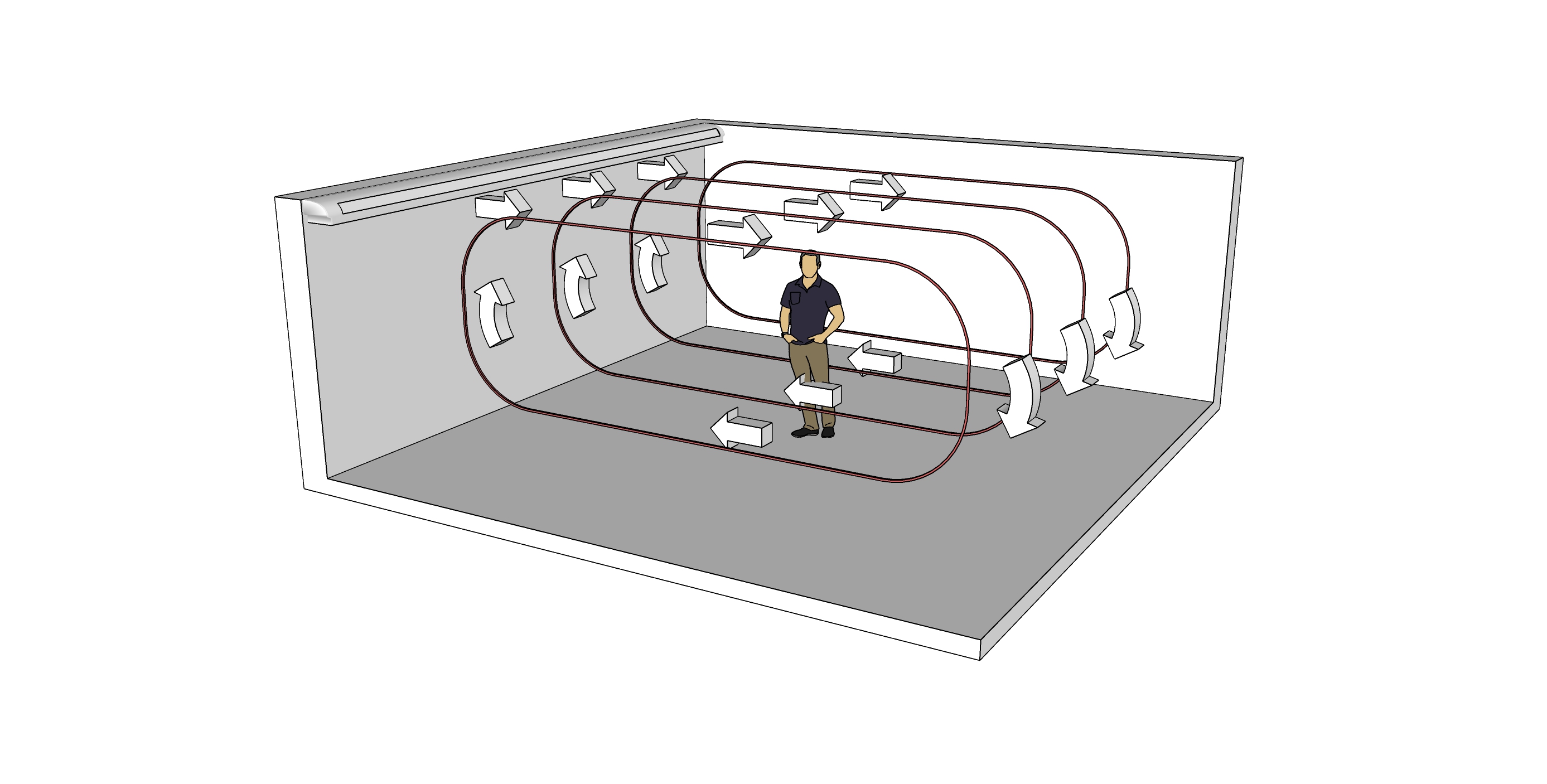}
    \end{overpic} 
    \vspace{-15mm}
\caption{Schematic of the room configuration illustrating the assumption of air recirculation in the room generated by an air-conditioning unit.}
    \label{fig:recirculation}
\end{figure}

Below we develop our model and explain major modelling assumptions. As discussed in the Introduction, while the WMR assumption of the Wells--Riley model leads to fast predictions, this does not capture any  effects due to the geometry. Since a CFD approach is very time consuming, we develop a quasi-3D model for the concentration of airborne viral particles indoors which provides rapid predictions. We assume an infectious person talking or breathing with or without a mask. We also assume that the room contains an air-conditioning unit, which drives a recirculating flow, as shown in Figure~\ref{fig:recirculation}, and as predicted by \cite{van2013suitability}. We then derive the concentration at all points on this recirculating loop. We will show that this model possesses a semi-analytic solution, which means that predictions can be generated very quickly. 


The infectious person acts as a source of infectious particles, $S_{\text{inf}}$. For simplicity, we make the assumption that all airborne particles are of the same size and carry the same amount of virus, but note that this can be readily generalised for particle size distributions. The airborne particles are transported by advection caused by the airflow which follows the recirculating loop, as shown in Figure~\ref{fig:recirculation}, which has velocity, $\vec{v}$. Even though air is expelled from the mouth and nose during breathing and talking, this is a local and temporary effect, so we assume for simplicity that the particles are passively released into the flow stream, so that the advection velocity of the particles is also $\vec{v}$.

The infectious particles are removed due to three factors: the ventilation system, biological deactivation of the virus, and gravitational settling of the virus. These correspond to sink terms, which we will denote by $S_{\text{vent}}$, $S_{\text{deact}}$, and $S_{\text{set}}$ respectively.

Finally, we assume turbulent mixing of the air. Turbulence leads to the  particles diffusing much more rapidly than due to Brownian motion~\cite{Nicas2009}. Turbulent diffusion is governed by the eddy diffusion coefficient, $K$ (m$^2$/s) \cite{Nicas2009}. 

From these assumptions, we arrive at the advection--reaction--diffusion equation, the governing equation for the concentration of infectious airborne particles: \begin{equation} \label{eq:adv-diff-eq} \frac{\partial \con}{\partial t} + \nabla \cdot (\vec{v} \con) - \nabla \cdot (K \nabla \con) = S_{\text{inf}} - S_{\text{vent}} - S_{\text{deact}} - S_{\text{set}}, \end{equation}
where $\con$ is the concentration of airborne infectious particles (particles/m$^2$) at all points in the surface of the looping airflow, $t$ is the time (s)~\cite{transport}, and  $\nabla=(\partial/\partial \xi, \partial/\partial y)$ is the two-dimensional gradient operator in the surface of the looping airflow, where $\xi$ (m) denotes the arclength in the streamwise direction of the flow and $y$ (m) is the coordinate transverse to the flow. The upper and lower branches of the surface are assumed to be separated by a distance $h/2$ \cite{van2013suitability}---see Figure~\ref{fig:layers+BC}(a).

\subsection{Further Modelling Assumptions--simplified ADR equation}

Following \cite{Khalid2020}, we model an infected person who is breathing or talking as a continuous point source emitting respiratory particles into the plane of the looping airflow at a constant rate of $R$ particles/s. We also assume that inhaling and exhaling occur at the same rate, so there is no net source or sink of air from the emitter and receiver of the particles. Thus, an infectious person talking or breathing at position ($\xi_0$,$y_0$) is modelled as follows: 
\begin{equation} \label{eq:source} S_{\textrm{inf}} = R \delta(\xi-\xi_0) \delta(y-y_0) \end{equation}
where $\delta(x)$ is the Kronecker delta function. Note that we can also model the infectious person vacating the room at a later time $t_1$ by adding a factor of $(1-H(t-t_1))$ to \eqref{eq:source}. (We consider this in Figure~\ref{fig:C-vs-t-lunch}.) 

We assume there is mechanical ventilation in this indoor space provided by air vents. Following \cite{gammaitoni1997using,Drivas1996}, we model the ventilation effect as a sink term of uniform strength over the domain,
\begin{equation} \label{eq:vent} S_{\textrm{vent}} = - \lambda \con, \end{equation}
where $\lambda$ is the air exchange rate of the room, measured in $\rm s^{-1}$. This is an approximation since removal of the air by ventilation occurs at a higher rate near the air vents \cite{Shao2017} but our approach still allows us to include the effect of ventilation while retaining its key features and without greatly increasing the mathematical complexity of our model.

Following \cite{gammaitoni1997using}, we model the biological deactivation and gravitational settling as global first-order removal:
\begin{align} 
\label{eq:deact-sink}
 S_{\textrm{deact}} &= -\beta \con,   \\
\label{eq:set-sink}
 S_{\textrm{set}} &= -\sigma \con,   
\end{align} 
where 
$\beta$ is the viral deactivation rate and $\sigma$ is the particle settling rate, all measured in $\rm s^{-1}$.

The typical airflow from an air-conditioning unit, $|\vec{v}|=0.1-1$\,m/s \cite{ASHRAE}, which is small compared to the speed of sound (340 $\rm{m/s}$), so we can assume that the air is an incompressible fluid \cite{Acheson}. 
\begin{equation} \label{eq:incompressible-fluid} \nabla \cdot \vec{v} = 0. \end{equation} 

We assume that $\vec{v}$ is solely controlled by the air-conditioning unit, which we assume has an inlet in the left wall, and we assume that the airflow velocity at all points in the surface of the loop is constant and uniform, ie. $\vec{v}=(v,0)$, where $v$ is a constant.

To determine $K$, the eddy diffusion coefficient, we use the following formula, which is valid for an isothermal room served by mixing ventilation. This was presented in \cite{Foat2020} and is based on the turbulent kinetic energy balance (TKEB) relationship initially proposed in \cite{Karlsson1994}:
\begin{equation} \label{eq:TKEB} K = c_v Q (2c_{\epsilon} V N^2)^{-1/3}. 
\end{equation}
Here, $c_v$ is the von Karman constant, $Q$ is the total volume flow rate into the room, $V$ is the room volume, $N$ is the number of air supply vents, and $c_{\epsilon}$ is the constant of proportionality in Taylor's Dissipation Law \cite{Karlsson1994,Taylor1935}. Since $Q = V\lambda$, \eqref{eq:TKEB} may be rewritten as
\begin{equation} \label{eq:K} K = c_v V\lambda (2c_{\epsilon} V N^2)^{-1/3}.  \end{equation}
We assume the relationship $c_{\epsilon} = c_v^3$ following \cite{Karlsson1994,Bodin1979}. Taking a value $c_v=0.39$ \cite{Andreas2009} gives $c_{\epsilon} = 0.059$. We note, however, that there is some uncertainty about the value of $c_{\epsilon}$. Specifically, some applications of the TKEB in the literature have used other formulae for $c_{\epsilon}$, for example $c_{\epsilon} = c_v$ \cite{Drivas1996} and $c_{\epsilon} = 0.5 c_v^3$ \cite{Foat2020}, without providing a rationale. 

Assuming there is only one infectious person in the room, we substitute equations  \eqref{eq:source}, \eqref{eq:vent}, \eqref{eq:deact-sink}, \eqref{eq:set-sink} and \eqref{eq:incompressible-fluid}  into the advection--diffusion--reaction equation \eqref{eq:adv-diff-eq} and obtain the partial differential equation (PDE) we will be solving:
\begin{equation} \label{eq:basePDE} \frac{\partial \con}{\partial t} + v\frac{\partial \con}{\partial \xi} - K \left(\frac{\partial^2 \con}{\partial \xi^2}+\frac{\partial^2 \con}{\partial y^2}\right) = R \delta(\xi-\xi_0) \delta(y-y_0) - (\lambda + \beta + \sigma) \con.  \end{equation}

We assume no virus-carrying aerosols initially. Hence, the initial condition is
\begin{equation}
    \label{eq:IC} \con(\xi,y,0) = 0.
\end{equation}

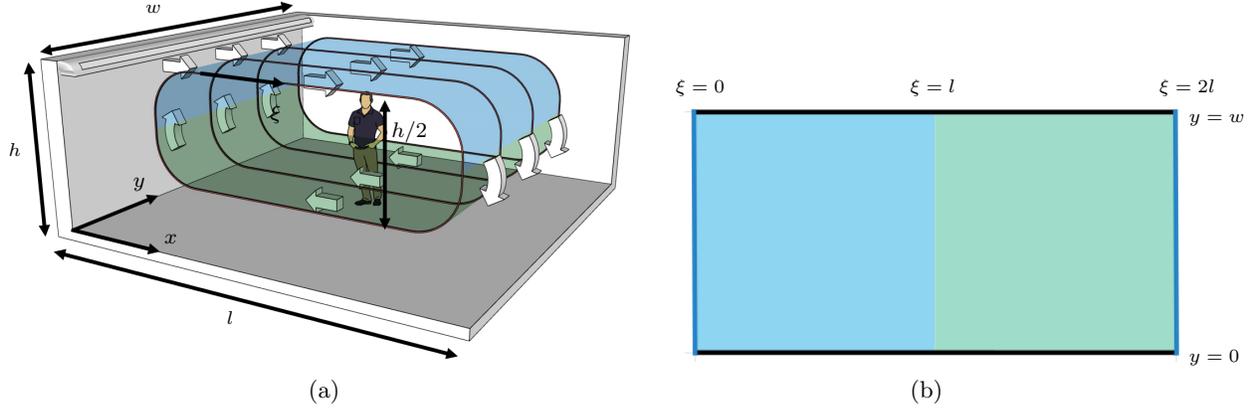
\begin{figure}[t]
\vspace{5mm}
      \subfloat[\label{fig:layers}]
   {
\begin{overpic}[width=.50\textwidth,tics=10]{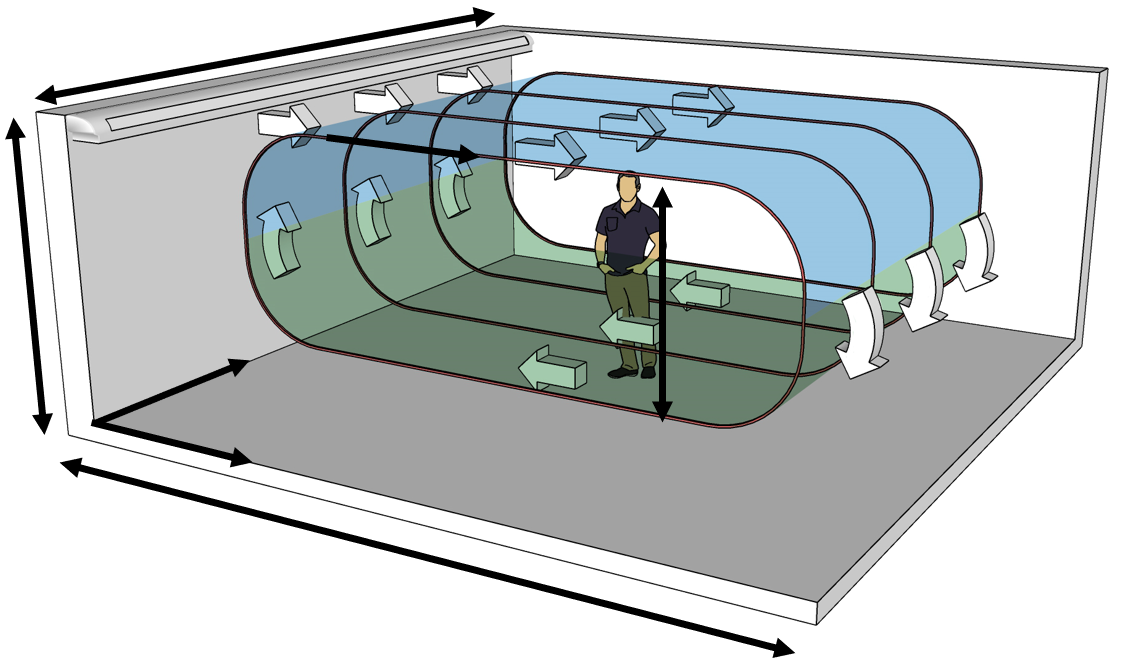}
\put(18,28){\footnotesize$y$}
\put(23,19){\footnotesize$x$}
\put(20,56){\scriptsize{$w$}}
\put(-2,33){\scriptsize{$h$}}
\put(33,6){\scriptsize{$l$}}
\put(40,39){\footnotesize$\xi$}
\put(59.5,36){\footnotesize{$h/2$}}
\end{overpic} }
\hspace{2mm}
\subfloat[\label{fig:BC}]{
\begin{overpic}[width=.40\textwidth,tics=10]{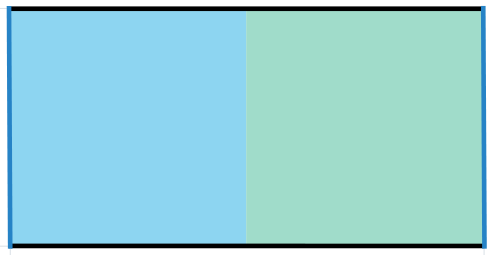}
\put(-2,54){\scriptsize{$\xi=0$}}
\put(45,54){\scriptsize$\xi=l$}
\put(95,54){\scriptsize$\xi=2l$}
\put(101,48){\scriptsize$y=w$}
\put(101,0){\scriptsize$y=0$}
\end{overpic} }
    \caption{(a) Quasi-3D setup from the assumption of air recirculation in a room of dimensions $l \times w \times h$; $\xi$ denotes the arclength around the loop surface of the recirculating flow. We divide the loop into an upper (blue/darker) and lower (green/lighter) domain; $x$ and $y$ denote Cartesian coordinates that describe the position on the floor. The upper half of the loop is projected onto the Cartesian plane via $x=\xi$, $0 \le \xi \le l$; the lower half is connected via $x=2l-\xi$, $l \le \xi \le 2l$. The upper and lower branches of the surface are assumed to be separated by a distance $h/2$.  (b) The unwrapping of the loop surface leads to a two-dimensional domain.
    }
    \label{fig:layers+BC}
\end{figure}

    We model the recirculating flow using the following domain and boundary conditions. For a room with length $l$ and width $w$, we unwrap the loop surface of the airflow to the two-dimensional domain $(\xi,y) \in [0,2l] \times [0,w]$ (see Figure~\ref{fig:layers+BC}(b)). This extended domain allows us to model the evolution of the aerosol cloud in both the upper and lower layers of the flow stream in a simpler way. In this setup, the aerosol cloud rejoins the original stream through periodic boundary conditions on the concentration and its derivative at the wall $\xi=0$ and opposite side of the domain at $\xi=2l$: 
    \begin{subequations} \label{eq:BC}
    \begin{align}
        \label{eq:x-BC1} & \con(0,y,t) = \con(2l,y,t) \\
        \label{eq:x-BC2} & \frac{\partial \con}{\partial \xi} (0,y,t) = \frac{\partial \con}{\partial \xi}(2l,y,t).
        \end{align}
    The model is closed by applying Neumann boundary conditions at the walls located at $y=0,w$:    
        \begin{align}
        \label{eq:y-BC} & \frac{\partial \con}{\partial y} (\xi,0,t) = \frac{\partial \con}{\partial y} (\xi,w,t) = 0, 
    \end{align}
    \end{subequations}
   as we assume that no particles pass through the other two walls.

\subsection{The semi-analytic solution: viral particle concentration}
\label{Section:Solution}
We solve the PDE \eqref{eq:basePDE} with the boundary conditions \eqref{eq:BC} and initial condition \eqref{eq:IC} by first solving the homogeneous problem to determine the impulse function. Then, we convolve the impulse with the source function \eqref{eq:source} to obtain the full solution \cite{Evans}. 

A solution to the homogeneous problem of the form $\con(\xi,y,t) = \mathrm{e}^{-(\lambda+\beta+\sigma) t} A(\xi,y,t)$ is assumed~\cite{Evans}. We then use separation of variables to reduce the problem to two one-dimensional diffusion problems~\cite{Evans}. Due to the periodic and Neumann boundary conditions \eqref{eq:BC}, we can use the method of images \cite{Carslaw-Jaeger} when solving for the impulse function. Hence, the solution is given by
\begin{equation} \label{eq:base-sol} \begin{split} \con(\xi,y,t) = &  \int_{0}^{t} \frac{R }{4\pi K\tau} \mathrm{e}^{-(\lambda+\beta+\sigma) \tau}  \sum_{m=-\infty}^{\infty} \mathrm{e}^{-\frac{(\xi - v \tau - \xi_0 - 2ml)^2}{4K\tau}}  \times \\
& \sum_{n=-\infty}^{\infty} \left( \mathrm{e}^{-\frac{(y - y_0 - 2nw)^2}{4K\tau}}    + \mathrm{e}^{-\frac{(y + y_0 - 2nw)^2}{4K\tau}}  \right) \mathrm{d}\tau. \end{split}  \end{equation}
Now, leveraging the mapping of the quasi-3D setup to a two-dimensional geometry we can average the particle concentration in the upper and lower branch of the loop surface to determine the concentration for the quasi-3D setup. Specifically, we define $x \in [0,l]$ to be the Cartesian coordinate in the direction of the upper flow, so $x=\xi$, $0 \le \xi \le l$ defines the upper branch and $x=2l-\xi$ defines the lower branch, for $l \le \xi \le 2l$ (see Figure~\ref{fig:layers+BC}(a)). The concentration of viral aerosols in the upper branch (see Figure~\ref{fig:layers+BC}(a)) is then given by 
\begin{equation}
    C_{\text{upper}}(x,y,t) = \con(x,y,t),
\end{equation}
and the concentration of viral aerosols in the lower branch is given by 
\begin{equation}
    C_{\text{lower}}(x,y,t) = \con(2l-x,y,t).
\end{equation}
Recall that we assumed that the separation between the two branches is half the height of the room. We assume that within these the two branches the air is well-mixed, and use this to transform the two-dimensional expressions back into a three-dimensional representation, 
\begin{equation} \label{eq:3d-total-conc} 
C(x,y,t) =  \frac{(C_{\text{upper}}(x,y,t)+C_{\text{lower}}(x,y,t))}{h/2}. 
\end{equation}
We then substitute \eqref{eq:base-sol} into \eqref{eq:3d-total-conc} to obtain the total concentration
\begin{equation} \label{eq:indoors-sol} 
 \begin{split} C(x,y,t)  
 = & \frac{2R }{4\pi Kh} \int_{0}^{t} \sum_{m=-\infty}^{\infty} \left( \mathrm{e}^{-\frac{(x - v \tau - x_0 - 2ml)^2}{4K\tau}} + \mathrm{e}^{-\frac{(x + v \tau + x_0 - 2ml)^2}{4K\tau}}  \right) \\
& \times\sum_{n=-\infty}^{\infty} \left( \mathrm{e}^{-\frac{(y - y_0 - 2nw)^2}{4K\tau}} + \mathrm{e}^{-\frac{(y + y_0 - 2nw)^2}{4K\tau}} \right) \frac{e^{-(\lambda+\beta+\sigma) \tau}}{\tau}  \mathrm{d}\tau. \end{split}  \end{equation}
We will use the latter expression for the concentration to generate our simulations. 

\subsection{The Probability of Infection (Infection Risk)}
We assume an exponential probability density function for infection as a function of the dose of viral particles inhaled, $d$ \cite{Riley1978,burridge2021,Watanabe2010}. Hence, the probability of infection (infection risk), $P$, is expressed as
\begin{equation} \label{eq:dose-response}
    P(d) = 1 - \text{exp}(-dI),
\end{equation}
where $I$ is a constant that depends on the infectiousness of the virus \cite{burridge2021,Watanabe2010}.

In our model, we use the following equation to calculate $d$, adapted from \cite{Riley1978,Vuorinen2020},
\begin{equation} \label{eq:dose} d(x,y,t) = \int_0^{t} \rho C(x,y,\tau) \mathrm{d}\tau, \end{equation}
where $\rho$ is the average breathing rate, ie.~the amount of air inhaled by a person per second and $C(x,y,t)$ is given by \eqref{eq:indoors-sol}. Substituting \eqref{eq:dose} into \eqref{eq:dose-response}, we have
\begin{equation} \label{eq:prob}
    P(x,y,t) = 1 - \text{exp}\left( -I \int_0^t \rho C (x,y,\tau) \mathrm{d}\tau \right).
\end{equation}

\subsection{Ventilation Settings}
\label{sec:ventset}
Ventilation is increasingly considered a very important intervention against COVID-19 \cite{burridge2021ventilation}. We will later consider four different ventilation settings, corresponding to four different values of $\lambda$, the air exchange rate, as follows:
\begin{itemize}
    \item Scenario 1: Very poor ventilation.
    \item Scenario 2: Poor ventilation.
    \item Scenario 3: A pre-pandemic ASHRAE recommended ventilation setting.
    \item Scenario 4: A pandemic-updated ASHRAE recommended ventilation setting, 
\end{itemize}
where ASHRAE is the American Society of Heating, Refrigeration and Air-Conditioning Engineers \cite{ASHRAE}.

\section{Parameters}
\label{Section:Parameters}

\subsection{Parameter Estimates}
\label{section:Parameter Estimates}

In this section we discuss the model parameters that are independent of the room configuration and give the values we take. In Section~\ref{section:Location-Specific Parameters} we address the remaining model parameters, which are location specific.

\subsubsection{Viral Particle Emission Rate (and Masks)}
Following \cite{Asadi2019} we assume that the emission rate of airborne particles, for breathing  $R=0.5$ particles/s and for talking $R=5$ particles/s.

We quantify the reduction in the infection risk when the infectious person wears a face mask. Therefore, we introduce a parameter, $\eta$, that measures the efficiency of the mask, so 
\begin{equation}
    R_{\textrm{mask}} = R (1-\eta).
\end{equation}
The efficiency of surgical masks and N95 respirators is well documented \cite{Li2006,Dbouk2020}. However, due to the sudden demand due to the COVID-19 pandemic, there has been a shortage of such masks and many people have had to manage with cloth masks. The efficiency of these masks is not well studied: at the moment there is only a preliminary study by \cite{Fischer2020}, which suggests that the efficiency is at least 50\%.
Furthermore, a fraction of people wear their masks incorrectly, which reduces their efficiency. Thus, we shall assume that a mask has an efficiency of 50\% which approximates a `bad-case' but probably realistic scenario.

We note that in \eqref{eq:indoors-sol}, $C$ is proportional to the strength of the source, $R$, as expected.  Henceforth, we will use the dimensionless parameter $\mathcal{R}=R/R_0$, where $R_0$ is the rate of emission of infectious particles when breathing (0.5\,particles/s, see Table~\ref{tab:parameters}). We work with $\mathcal{R}$ in all subsequent analysis (see Table~\ref{tab:parametersfor R}).
\begin{table}[h]
\centering
\captionof{table}{The rate of emission of infectious particles by a person and the corresponding non-dimensional rate $\mathcal R$ when breathing or talking with and without a mask, scaled by the emission rate during breathing, $R_0=0.5$\,particles/s.}
\begin{tabular}{|c|c|c|}
\hline
Infectious person state & $R$ (particles/s) & $\mathcal{R}=R/R_0$  \\
\hline 
Breathing & 0.5 & 1 \\
Talking & 5 & 10 \\
Breathing with mask & 0.25 & 0.5\\ 
Talking with mask & 2.5 & 5 \\
\hline
\end{tabular}

\label{tab:parametersfor R}
\end{table}

\subsubsection{Viral Deactivation and Gravitational Settling}
For the viral deactivation rate, $\beta$, according to \cite{Doremalen2020} artificially generated aerosols carrying SARS-CoV-2 had a half-life of 1.1--1.2 hours which suggests that $\beta$ is in the range of 0.58--0.63h$^{-1}$ $\approx$ 1.6--1.8 $\times 10^{-4}$~s$^{-1}$. In our simulations, we will take $\beta=1.7\times 10^{-4}$~s$^{-1}$. We must caution that this value of $\beta$ is for artificially generated aerosols in a controlled laboratory environment and may be different for naturally produced bio-aerosols. However, our model can easily be updated if more accurate estimates can be obtained.


For the gravitational settling parameter, $\sigma$, we refer to \cite{Oliveira2021}, in which aerosol droplets released by respiratory activities were studied. This gave a steady-state mean value of $\sigma = 0.39$~h$^{-1}$ $\approx$ 1.1$\times 10^{-4}$~s$^{-1}$, which we will use in our work. 

\subsubsection{Infectiousness Constant}
The final parameter we need to calculate the probability of infection in \eqref{eq:prob} is $I$. This parameter is related to the median infectious dose, $d_m$, defined as the dose that causes infection in 50\% of the population, by
\begin{align}
\label{eq:k dm relation}
     0.5 = 1 - \exp(-d_m I).
\end{align}
At present, $d_m$ is unknown for SARS-CoV-2. An early estimate by \cite{Vuorinen2020}, however, suggests that $d_m$ for airborne COVID-19 is $100$ inhaled particles. 
In \cite{Franz1997} a list of the infectious dose of several biological warfare agents are presented, which show that bacteria and virus aerosols can cause disease with as few as 1--100 aerosols in two hours. Therefore, this estimate for SARS-CoV-2 of $100$ particles is reasonable, keeping in mind that it may have to be updated in the future. Using this value of $d_m$, \eqref{eq:k dm relation} gives $I\approx 0.0069$. 

We collect values for the parameters discussed in this section in Table~\ref{tab:parameters}. The results of our model should be interpreted with caution until these values are confirmed.

\begin{table}[t]
\centering
\captionof{table}{Parameter estimates }
\begin{tabular}{|p{1.1in}|c|p{2.1in}|c|}
\hline
Parameter & Symbol & Value & Source \\
\hline
\multirow{2}{1.5in}{Generation rate of infectious particles} 
& $R$ & Breathing: 0.5 particle/s & \cite{Asadi2019} \\
& & Talking: 5 particles/s & \cite{Asadi2019} \\
\hline
Efficiency of mask & $\eta$ & 0.5 & \cite{Fischer2020} \\ \hline
  \multirow{2}{1in}{Room air exchange rate} & $\lambda$ & Very poor ventilation: 0.12 h$^{-1} \approx 3.3 \times 10^{-5}$ s$^{-1}$ & \cite{Guo2008}\\
  & & Poor ventilation: 0.72 h$^{-1} \approx 2 \times 10^{-4}$ s$^{-1}$ & \cite{Guo2008} \\
  & & Pre-pandemic recommended ventilation: 3 h$^{-1} \approx 8.3 \times 10^{-4}$ s$^{-1}$ & \cite{ASHRAE-rec} \\
& & Pandemic-updated recommended ventilation: 6 h$^{-1} \approx 1.7 \times 10^{-3}$ s$^{-1}$ & \cite{web:ASHRAE-rec2} \\ \hline
Virus deactivation rate & $\beta$ & 1.7$\times 10^{-4}$~s$^{-1}$ & \cite{Doremalen2020} \\ \hline
Aerosol settling rate & $\sigma$ & 1.1$\times 10^{-4}$~s$^{-1}$ & \cite{Oliveira2021}\\ 
\hline
Median infectious dose & $d_m$ & 100 particles  & \cite{Vuorinen2020} \\
\hline
\end{tabular}

\label{tab:parameters}
\end{table}

\subsection{Ease of Updating the Model}
\label{Section:Ease-Update}

The results of our model can easily be updated when the value of $R$ is confirmed. As $R$ is dependent on the breathing rate of the infectious person \cite{Asadi2019} and the efficiency of masks \cite{Li2006,Fischer2020}, we can also easily update our results for different scenarios, such as a person wearing a surgical mask or a person exercising in a gym.
\subsubsection{Updating the Infectiousness Constant}
Regarding the probability of infection in \eqref{eq:prob}, a key uncertainty currently about SARS-CoV-2 is the infectiouness constant $I$. Our model can be easily updated for a confirmed value of $I$ by making use of the following observation. Suppose our model uses a value $I=I_0$ and obtains a result $P_0$ for the probability of infection. Then, if  we are given an updated value for the infectiousness constant, $I=I_1$, the new probability of infection, $P_1$ is given by
\begin{equation} \label{eq:prob-scale} P_1(x, y, t) = 1 - (1-P_0)^{I_1/I_0}. \end{equation}

\subsubsection{The Infection Risk for New Virus Variants}
Expression \eqref{eq:prob-scale} can be used to quickly predict the infection risk for \emph{new variants} of SARS-CoV-2 that are being discovered. Since the virus is very small compared with the size of an aerosol particle in which it is transported \cite{Jin2020}, we can safely assume that the new variants do not affect the transport of the infectious aerosols in the air. Hence, we can assume that the concentration \eqref{eq:indoors-sol} still holds. The main difference between the virus variants in our model is the rate of virus-carrying aerosols generated, $R$, and the infectiousness of the variant, which is described by $I$. Both of these can be easily updated in our model through the steps discussed previously.

\subsubsection{Changing the Breathing Rate}
In a similar manner, we can update the infection risk if the susceptible person has, for example, a \emph{higher breathing rate through physical activity}. Suppose our model uses a value $\rho=\rho_0$ and obtains a result $P_0$ for the probability of infection. Then, if  we are given a new value for the breathing rate, $\rho=\rho_1$, the new probability of infection, $P_1$ is given by
\begin{equation} \label{eq:prob-scale-rho} P_1(x, y, t) = 1 - (1-P_0)^{\rho_1/\rho_0}. \end{equation} 
\subsection{Low-cost Numerical Implementation}
In the subsequent simulations, we implement the solution for the concentration \eqref{eq:indoors-sol} in Python 3.8.5 64-bit. The convolutions are  performed using the \textit{convolve} function from the scipy.signal subpackage, which convolves two $N$-dimensional arrays~\cite{2020SciPy-NMeth}. To achieve satisfactory accuracy, we use a time step of 1s.
The first infinite series in \eqref{eq:indoors-sol} is evaluated for $m \le |vt / 2l|$ as this is the number of times a particle travels around the recirculation surface during the time $t$. The second infinite series in \eqref{eq:indoors-sol} needs to be evaluated for $-3\le n \le 3$, as the steep exponential decay of the terms ensures very good accuracy with only the first few terms in the series \cite{Shao2017}.

To determine the infection risk \eqref{eq:dose}, we apply the \textit{cumtrapz} function from the \linebreak scipy.integrate subpackage, which cumulatively integrates an array of values using the trapezoidal rule \cite{2020SciPy-NMeth}, to the results of the convolutions. All plots are produced using the Matplotlib library \cite{matplotlib}.

The contour plots of the concentration and probability of infection are generated using a rectangular mesh of size 0.05m. The computation time required is approximately one minute for one hour of real time.
We run the simulations on a 2012 MacBook Pro laptop, with a 2.5GHz Dual Core Intel Core i5-3210M processor and 4GB 1805MHz RAM. Our code is available at:\\\url{https://github.com/zechlau14/Modelling-Airborne-Transmission}.

\section{Results and Discussion}
\label{Section:Results and Discussion}

\subsection{Case Study: A classroom}
\label{Case Study: A classroom}
\subsubsection{Location-Specific Parameters}
\label{section:Location-Specific Parameters}

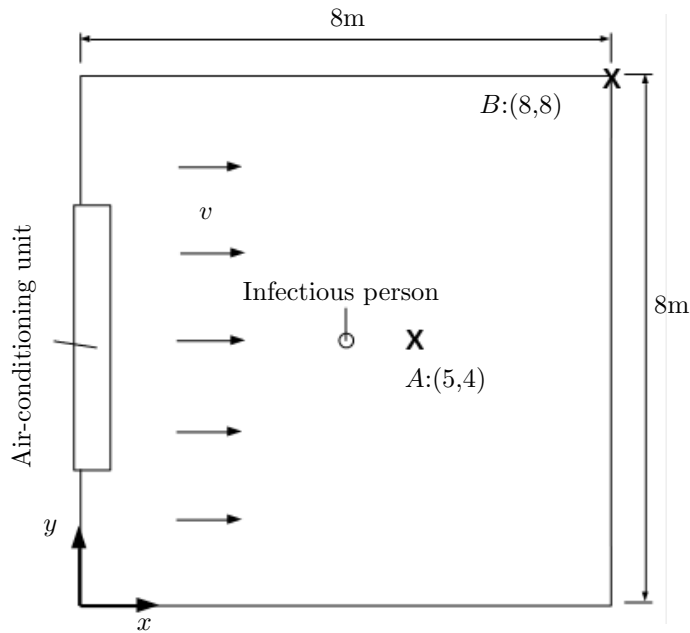
\begin{figure}[hb]
\begin{center}
\begin{overpic}[width=.5\textwidth,tics=10]{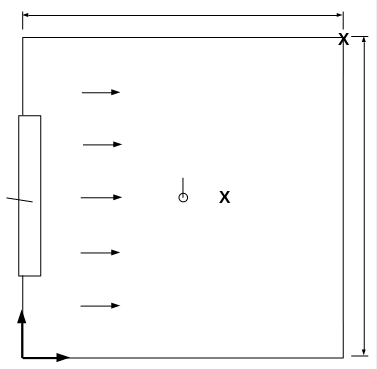}
\put(46,96){8m}
\put(98,50){8m}
\put(25,65){$v$}
\put(32,52){Infectious person}
\put(58,38){$A$:(5,4)}
\put(70,82){$B$:(8,8)}
\put(-5,25){\rotatebox{90}{Air-conditioning unit} }
\put(0,15){$y$}
\put(15,-1){$x$}
\end{overpic}
\caption{Schematic of the modelled classroom. One infectious person (viral source) is located at the centre of the room. Positions $A:(5,4)$ and $B:(8,8)$ are of particular interest and are studied in our analysis. }
\label{fig:schematic}
\end{center}
\end{figure}

Countries have been debating the best practices for operating indoor spaces in schools and universities during the COVID-19 pandemic. Hence, we decided to use our model to study the transmission of COVID-19 in an average classroom of dimensions 8m ($l$) $\times$ 8m ($w$) $\times$ 3m ($h$) -- see Figure \ref{fig:schematic}. All parameters associated with this location and their values are found in Table~\ref{tab:local-parameters}.

For this classroom, we assume that the air-conditioning drives an air flow with velocity $v=0.15$m/s \cite{ASHRAE}. We consider four physically relevant ventilation settings. For the very poor ventilation scenario (scenario $1$), we take $\lambda$ to be $0.12\rm{h}^{-1}$, the mean value determined for a closed classroom (closed windows, fans off) with the air-conditioning switched off, in a primary-school study by~\cite{Guo2008}. The study investigated the impact of air exchange rates on the indoor pollution particle number concentration and showed that, in practice, classrooms may have far lower air exchange rates than those recommended. For the poor ventilation scenario (scenario 2), we take $\lambda$ to be $0.72\rm{h}^{-1}$, the mean value determined for a closed classroom but now with the air-conditioning switched on \cite{Guo2008}.
For scenarios $3$ and $4$, we consider the recommended air exchange rate for classrooms made by the American Society of Heating, Refrigeration and Air-Conditioning Engineers (ASHRAE) before the pandemic, $3\rm{h}^{-1}$ \cite{ASHRAE-rec} and the recommendation they issued for the pandemic, $6\rm{h}^{-1}$~\cite{web:ASHRAE-rec2}, respectively. The aforementioned four values of $\lambda$ correspond to values of the eddy diffusion coefficient, $K$, respectively. $K$ is determined by \eqref{eq:TKEB} assuming $N=1$ and its values are given in Table \ref{tab:local-parameters}.

We also assume that a susceptible person inside a classroom is performing a sedentary activity (breathing or talking). During sedentary activity, the average person inhales $16$ times per minute with an air intake volume of approximately $0.5\rm L$ \cite{physio}. Hence, the breathing rate, $\rho$ is approximately $1.3 \times 10^{-4}$ m$^3$/s \cite{physio}.

\begin{table}[t]
\centering
\captionof{table}{Location-dependent parameters and their values}
\begin{tabular}{|p{1in}|c|p{2.2in}|c|}
\hline
Parameter & Symbol & Value & Source \\
\hline
Length of room& $l$ & 8m & \\
\hline
Width of room & $w$ & 8m & \\
\hline
Height of room & $h$ & 3m & \\
\hline
Airflow speed & $v$ & 0.15 m/s & \cite{ASHRAE} \\
\hline
  \multirow{2}{1in}{Eddy diffusion coefficient} & $K$ & Very poor ventilation: 8.8 $\times 10^{-4}$ m$^2$/s & \cite{Foat2020}  \\
 & & Poor ventilation: 5.3 $\times 10^{-3}$ m$^2$/s & \cite{Foat2020} \\
 & & Pre-pandemic recommended ventilation: 2.2 $\times 10^{-2}$ m$^2$/s & \cite{Foat2020} \\
 & & Pandemic-updated recommended ventilation: 4.4 $\times 10^{-2}$ m$^2$/s & \cite{Foat2020} \\
 \hline
Breathing rate & $\rho$ & $1.3 \times 10^{-4}$ m$^3$/s & \cite{physio} \\ \hline
\end{tabular}
\label{tab:local-parameters}
\end{table}

\subsubsection{The Concentration of Airborne Infectious Particles}
\label{Section: Concentration of Airborne Infectious Particles}
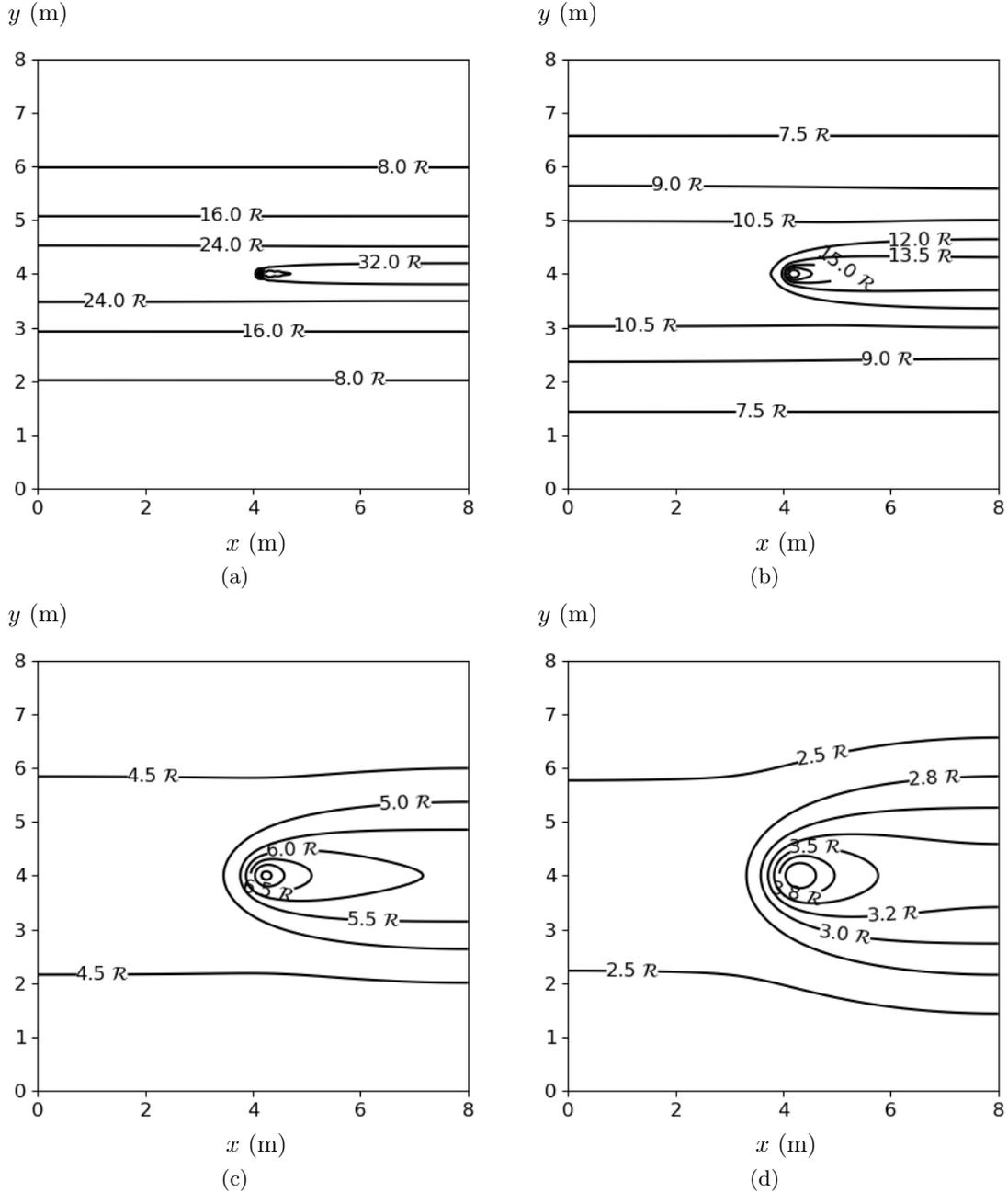
\begin{figure}[p]
\centering
\subfloat[\label{fig:C-contour-Q0}]{
\begin{overpic}[trim = 2.5cm 0 2.5cm 0, clip,width=.45\textwidth,tics=10]{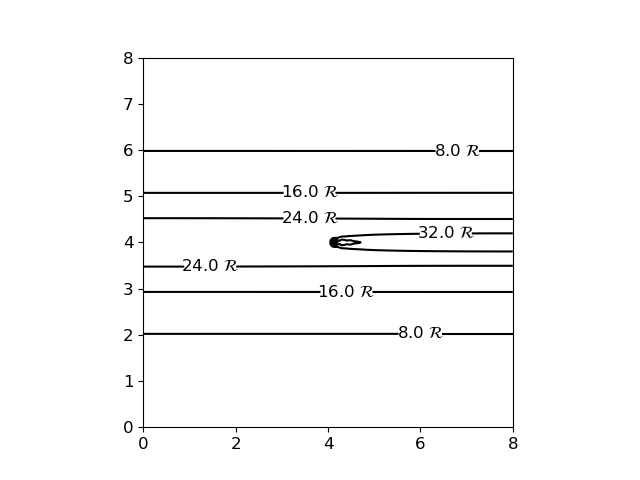}
\put(43,0){$x$ (m) }
\put(4,95){$y$ (m)}
\end{overpic} }
\subfloat[\label{fig:C-contour-Q1}]{
\begin{overpic}[trim = 2.5cm 0 2.5cm 0, clip,width=.45\textwidth,tics=10]{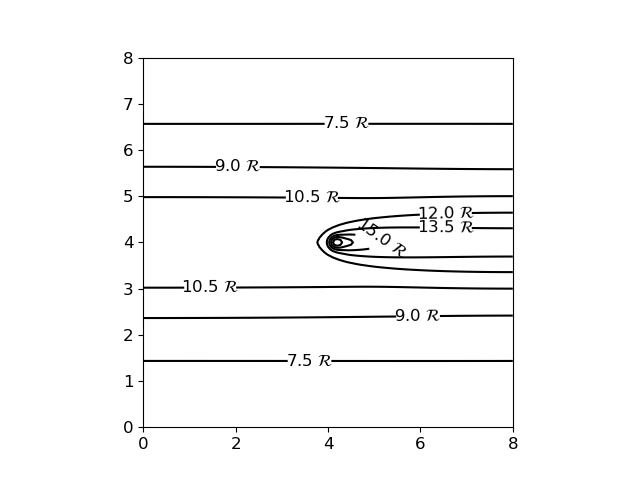}
\put(43,0){$x$ (m) }
\put(4,95){$y$ (m)}
\end{overpic} } \\
\subfloat[\label{fig:C-contour-Q2}]{
\begin{overpic}[trim = 2.5cm 0 2.5cm 0, clip,width=.45\textwidth,tics=10]{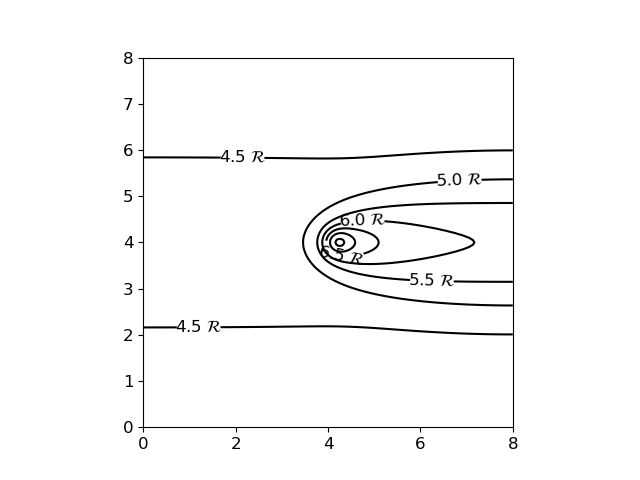}
\put(43,0){$x$ (m) }
\put(4,95){$y$ (m)}
\end{overpic} }
\subfloat[\label{fig:C-contour-Q3}]{
\begin{overpic}[trim = 2.5cm 0 2.5cm 0, clip,width=.45\textwidth,tics=10]{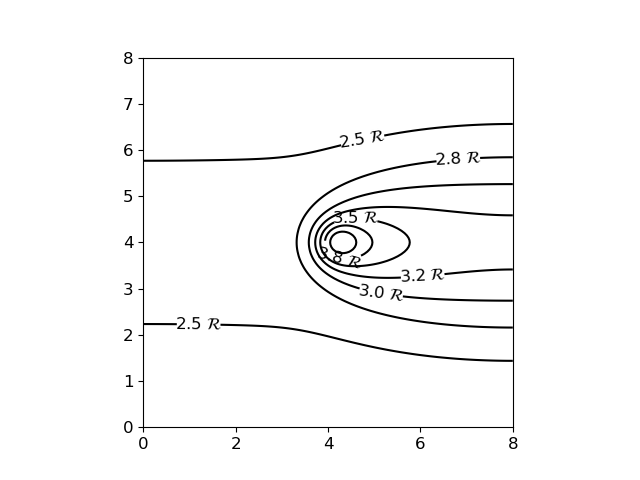}
\put(43,0){$x$ (m) }
\put(4,95){$y$ (m)}
\end{overpic} }
\caption{Concentration of SARS-CoV-2-carrying particles after one hour in an 8m $\times$ 8m $\times$ 3m room, determined from Equation \eqref{eq:indoors-sol}, with parameter values given in Tables \ref{tab:parameters} and \ref{tab:local-parameters}: (a) very poor ventilation, (b) poor ventilation, (c) pre-pandemic recommended ventilation, (d) pandemic-updated recommended ventilation. Values for $\mathcal{R}$ for breathing and talking with and without masks are given in Table~\ref{tab:parametersfor R}.}
\label{fig:C-contour}
\end{figure}

Figure \ref{fig:C-contour} shows the concentration of the infectious particles in the classroom after one hour. For any ventilation scenario, the highest concentration is in the region directly downwind from the infectious person. The width of this region increases with the amount of ventilation, as it depends on the eddy diffusion coefficient, $K$, which is proportional to $\lambda$. The next highest concentration in the room is found upwind, then the concentration decreases as one travels away from the infectious person in the direction orthogonal to the air flow. Our results agree qualitatively with those obtained from air sampling in hospital wards in Wuhan, conducted by \cite{Guo2020}, which showed that virus-carrying particles were `mainly concentrated near and downstream from the patients' and there was also an `exposure risk upstream'.

\begin{figure}
\centering
\subfloat[\label{fig:C-vs-t-XY54}]{
\begin{overpic}[trim = 1cm 0 1cm 0, clip,width=.45\textwidth,tics=10]{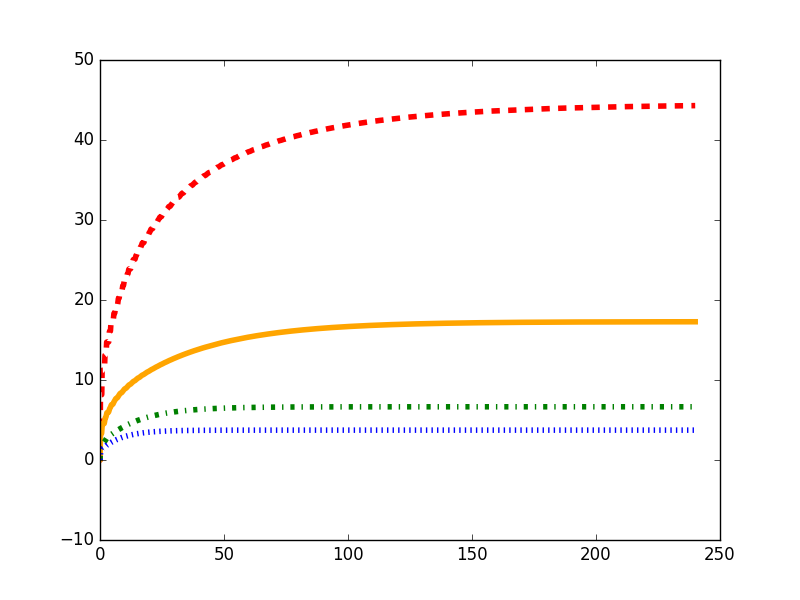}
\put(40,-2){Time (min) }
\put(-6,5){\rotatebox{90}{$C(x_A,y_A,t)$ ($\times \mathcal{R}$ particles/m$^3$)}}
\end{overpic} }
\subfloat[\label{fig:C-vs-t-XY88}]{
\begin{overpic}[trim = 1cm 0 1cm 0, clip,width=.45\textwidth,tics=10]{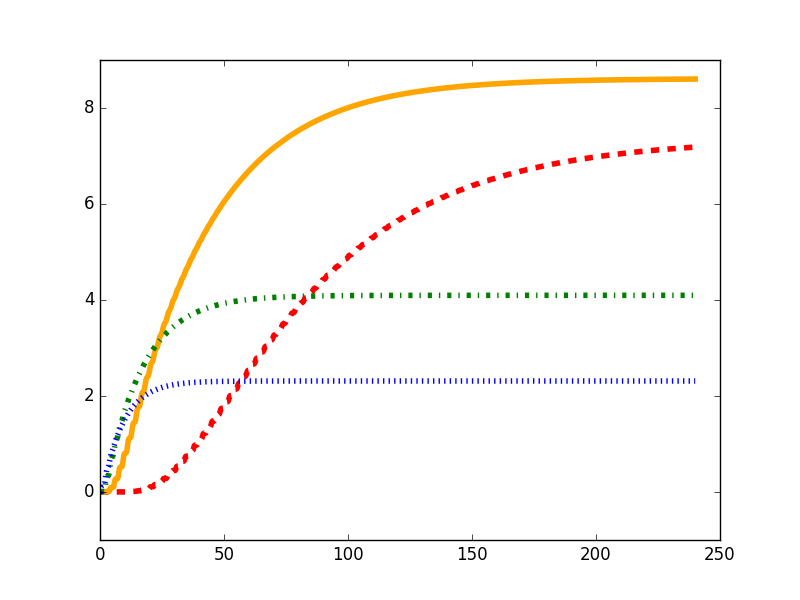}
\put(40,-2){Time (min) }
\put(-6,5){\rotatebox{90}{$C(x_B,y_B,t)$ ($\times \mathcal{R}$ particles/m$^3$)}}
\end{overpic} } 
\caption{Concentration of SARS-CoV-2-carrying particles versus time in a case of very poor ventilation (red dashed), poor ventilation (orange solid), pre-pandemic recommended ventilation (green dot-dashed) and pandemic-updated recommended ventilation (blue dotted), from Equation \eqref{eq:indoors-sol}, with parameter values as given in Tables \ref{tab:parameters} and \ref{tab:local-parameters}. (a) Evaluated at Position $A$:(5,4). In this case, a power law of the form $C(x_A, y_A, t) \propto \mathcal{R}t^{\alpha}$ is obeyed, where $\alpha \approx 0.34$; the trend deviates from this power law when $t \approx 16$ minutes and the presence of the walls begins to influence the solution.  (b) Evaluated at Position $B$:(8,8). In this case, the influence from the walls is significant for all times and hence no scaling law is obeyed. Values for $\mathcal{R}$ for breathing and talking with and without masks are given in Table~\ref{tab:parametersfor R}.}
\label{fig:C-vs-t}
\end{figure}

Figure \ref{fig:C-vs-t} shows the concentration in the room versus time evaluated at Position $A$:(5,4), and Position $B$:(8,8) (see Figure \ref{fig:schematic}). These two positions were chosen because, as seen in Figure \ref{fig:C-contour}, Position $A$ is where the highest concentration is while maintaining 1m social distancing from the infectious person and Position $B$ is where the lowest downstream concentration is. 
Figure \ref{fig:C-vs-t} shows that the concentration increases initially, before reaching a steady state.
For lower values of $\lambda$, more time is required to approach the steady-state concentration, as predicted also by the Gammaitoni--Nucci extension of the Wells--Riley model \cite{gammaitoni1997using}. 
At point $A$, downwind from the source, Figure~\ref{fig:C-vs-t-XY54} shows that the steady-state concentration is higher for lower values of $\lambda$, where there is less removal of infectious particles, as expected. 
Figure \ref{fig:C-vs-t-XY54} also shows that we can obtain a power law of the form $C(x_A,y_a,t) \propto \mathcal{R}t^{\alpha}$, where $\alpha \approx 0.34$ which is independent of the amount of ventilation; only the constant of proportionality depends on $\lambda$. The trend deviates from this power law (by more than 5$\%$) when $t \approx 16$ minutes.

For positions that are outside the high-concentration regions, such as Position $B$ in the corner, Figure \ref{fig:C-vs-t-XY88} shows that no scaling law is obeyed, due to the influence of the walls. Moreover, we see that the better ventilation scenarios 3 and 4 initially have higher concentrations before being overtaken by the worse ventilation scenarios. This result occurs because the movement of the infectious particles in the direction orthogonal to the airflow is governed by the eddy diffusion coefficient, $K$, which increases with $\lambda$ according to \eqref{eq:K}. Since the better-ventilation scenarios have higher values of $K$, the infectious particles diffuse faster in the $y$-direction to begin with. However, given sufficient time, the infectious particles in the worse ventilation scenarios will also reach the farthest points of the room, and the concentration there will eventually surpass that in the better-ventilation scenarios. This result is a first indication that a small amount of ventilation could actually increase the risk of transmission compared to the case of very poor ventilation.

\begin{figure}[p]
\centering
\subfloat[\label{fig:C-vs-t-XY54-lunch}]{
\begin{overpic}[trim = 1cm 0 1cm 0, clip,width=.45\textwidth,tics=10]{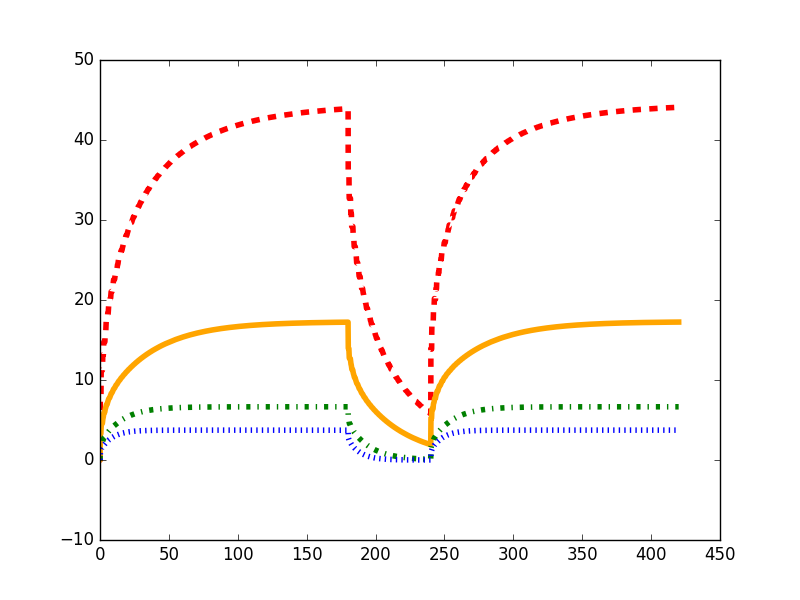}
\put(35,-2){Time (min) }
\put(-6,5){\rotatebox{90}{$C(x_A,y_A,t)$ ($\times \mathcal{R}$ particles/m$^3$)}}
\end{overpic} }
\subfloat[\label{fig:C-vs-t-XY88-lunch}]{
\begin{overpic}[trim = 1cm 0 1cm 0, clip,width=.45\textwidth,tics=10]{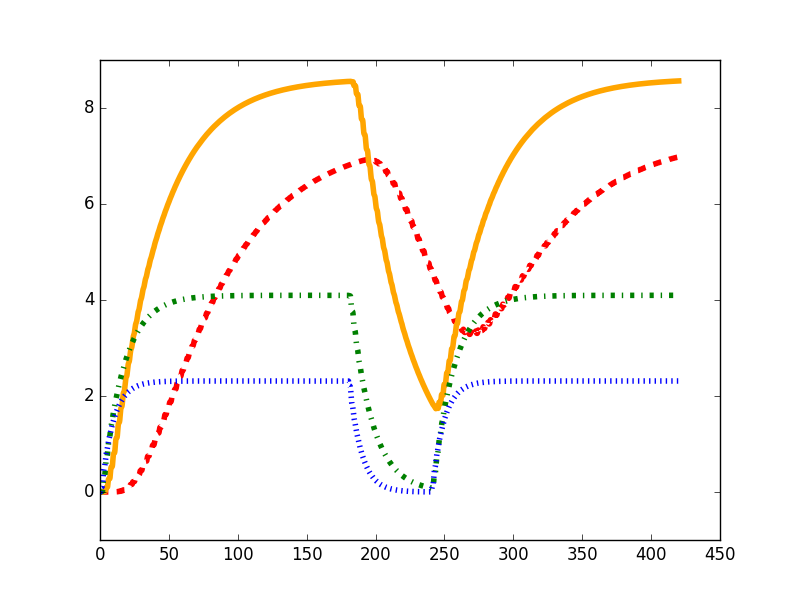}
\put(35,-2){Time (min) }
\put(-6,5){\rotatebox{90}{$C(x_B,y_B,t)$ ($\times \mathcal{R}$ particles/m$^3$)}}
\end{overpic} } 
\caption{Concentration of SARS-CoV-2-carrying particles versus time in a case of very poor ventilation (red dashed), poor ventilation (orange solid), pre-pandemic recommended ventilation (green dot-dashed) and pandemic-updated recommended ventilation (blue dotted) for a seven-hour school day with a one-hour lunch break, determined from solving \eqref{eq:basePDE} with source \eqref{eq:source}, set to zero for the lunch hour, and parameter values as given in Tables \ref{tab:parameters} and \ref{tab:local-parameters}. (a) Evaluated at Position $A$:(5,4).  (b) Evaluated at Position $B$:(8,8). Values for $\mathcal{R}$ for breathing and talking with and without masks are given in Table~\ref{tab:parametersfor R}.}
\label{fig:C-vs-t-lunch}
\end{figure}

Our model can also be used to find the vacancy time required in classroom settings. In Figure~\ref{fig:C-vs-t-lunch}, we present the concentration at Position $A$ and Position $B$ in the room for a seven-hour school day with a one-hour lunch break in the middle of the day. For the ASHRAE-recommended scenarios 3 and 4, the one-hour vacancy of the room is shown to be sufficient to reset the concentration in the room. However, for the poor and very poor ventilation scenarios 1 and 2, an hour is not enough to clear the room, and the concentration in the room will reach the steady-state concentration more quickly. 

\begin{figure}[p]
\centering
\subfloat[\label{fig:prob-Q0}]{
\begin{overpic}[trim = 2.5cm 0 2.5cm 0, clip,width=.45\textwidth,tics=10]{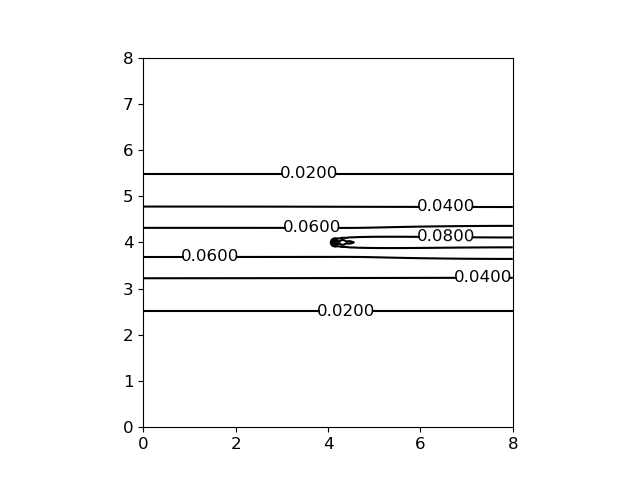}
\put(43,0){$x$ (m) }
\put(4,95){$y$ (m)}
\end{overpic} }
\subfloat[\label{fig:prob-Q1}]{
\begin{overpic}[trim = 2.5cm 0 2.5cm 0, clip,width=.45\textwidth,tics=10]{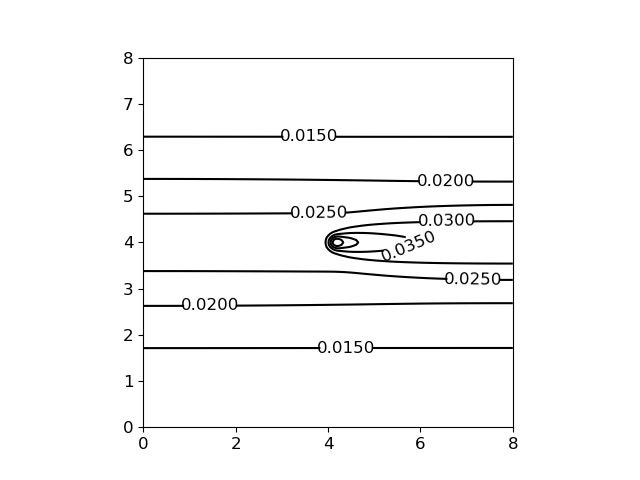}
\put(43,0){$x$ (m) }
\put(4,95){$y$ (m)}
\end{overpic} } \\
\subfloat[\label{fig:prob-Q2}]{
\begin{overpic}[trim = 2.5cm 0 2.5cm 0, clip,width=.45\textwidth,tics=10]{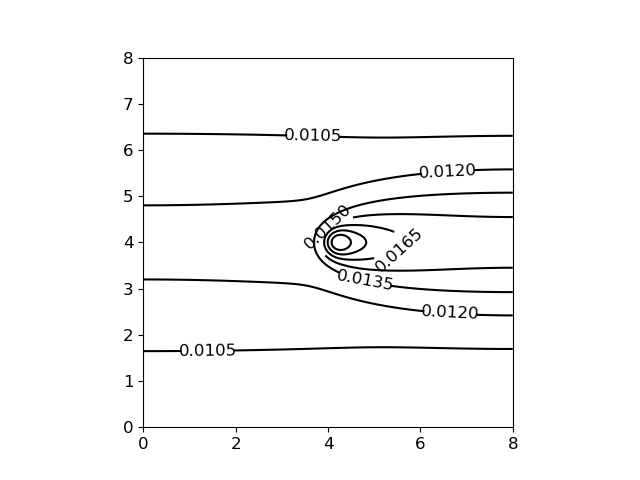}
\put(43,0){$x$ (m) }
\put(4,95){$y$ (m)}
\end{overpic} }
\subfloat[\label{fig:prob-Q3}]{
\begin{overpic}[trim = 2.5cm 0 2.5cm 0, clip,width=.45\textwidth,tics=10]{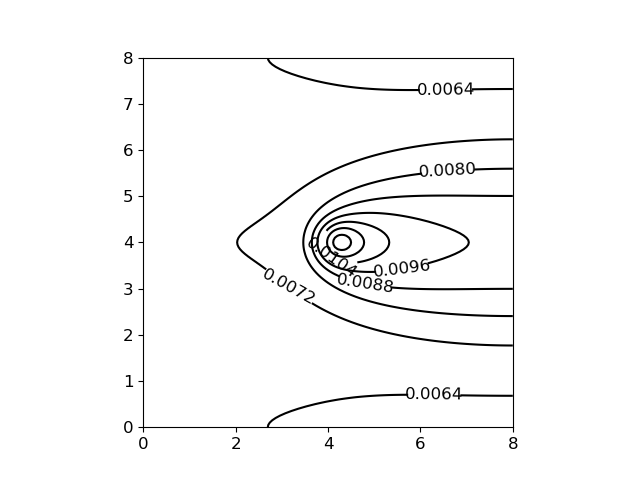}
\put(43,0){$x$ (m) }
\put(4,95){$y$ (m)}
\end{overpic} }
\caption{Probability of infection due to an infectious person breathing at the centre of a 8m $\times$ 8m $\times$ 3m room from \eqref{eq:indoors-sol} and \eqref{eq:prob}, with parameter values as given in Tables \ref{tab:parameters} and \ref{tab:local-parameters}: (a) very poor ventilation, (b) poor ventilation (c) pre-pandemic recommended ventilation (d) pandemic-updated recommended ventilation. Here, $\mathcal{R}=1$ which corresponds to breathing.} 
\label{fig:Prob-contour}
\end{figure}
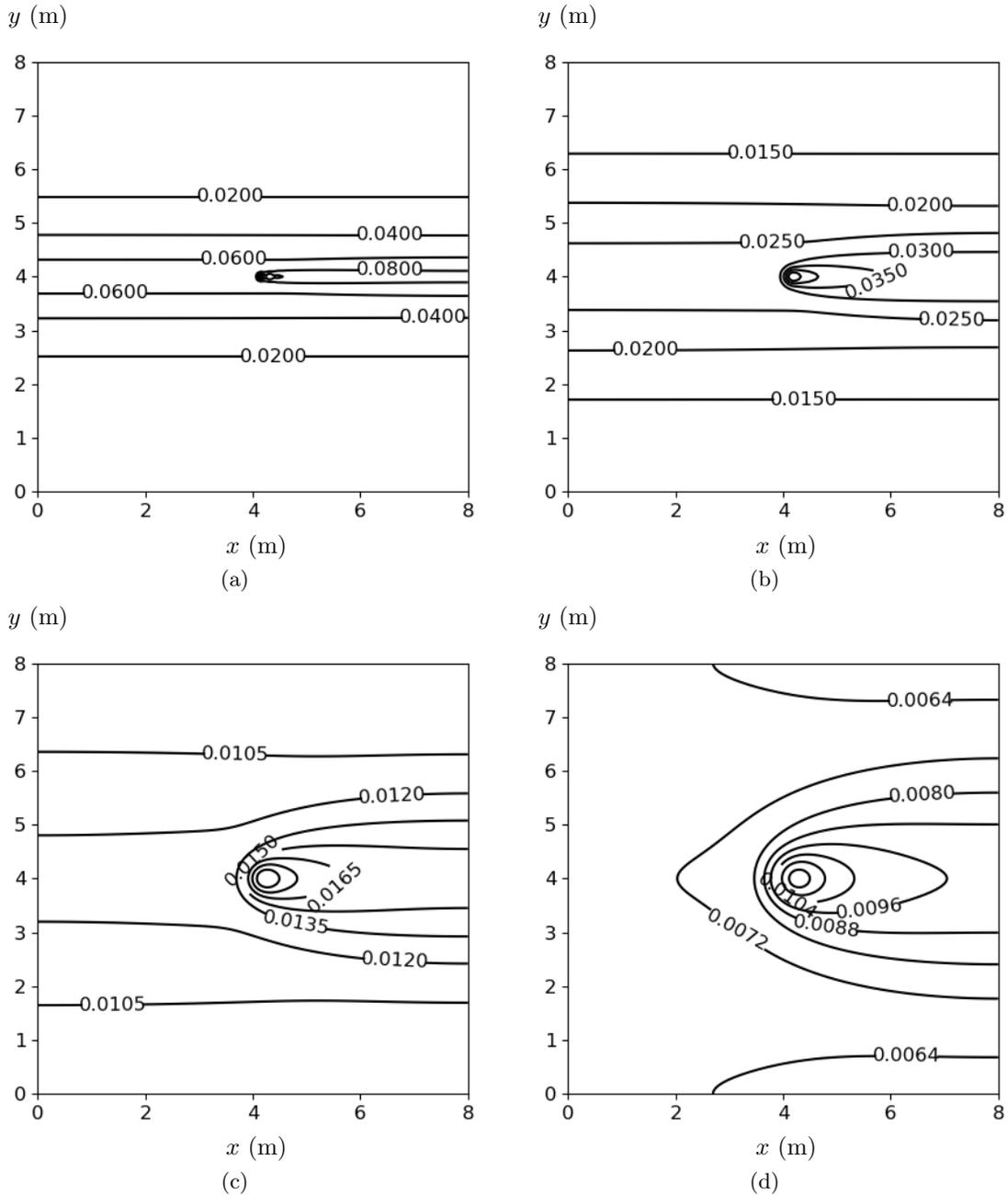

\newpage
\subsection{Probability of Infection}
\label{Section: Probability of airborne transmission}

Figure \ref{fig:Prob-contour} shows the probability of infection for the four ventilation settings we considered in the 8m $\times$ 8m $\times$ 3m room after an infectious person just breathes for one hour. The contours in Figure~\ref{fig:Prob-contour} are similar in shape to the concentration contours in Figure~\ref{fig:C-contour}. 
This implies that the greatest risk of infection indoors is directly downwind from the infectious person, and the risk decreases as we travel away from the source in a direction orthogonal to the airflow.

\begin{figure}[h]
\centering
\subfloat[\label{fig:Prob-vs-t-XY54}]{
\begin{overpic}[trim = 1cm 0 1cm 0, clip,width=.45\textwidth,tics=10]{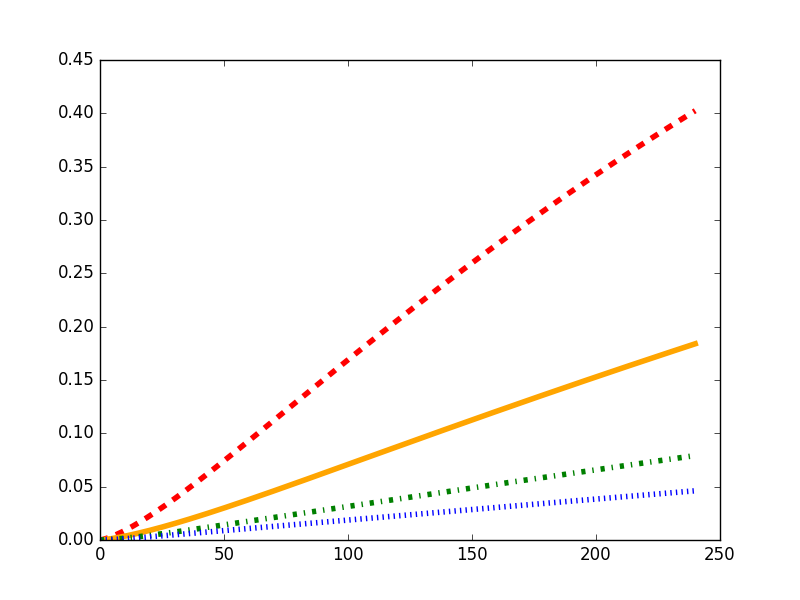}
\put(34,-2){Time (min) }
\put(-6,20){\rotatebox{90}{$P(x_A,y_A,t)$}}
\end{overpic} }
\subfloat[\label{fig:Prob-vs-t-XY88}]{
\begin{overpic}[trim = 1cm 0 1cm 0, clip,width=.45\textwidth,tics=10]{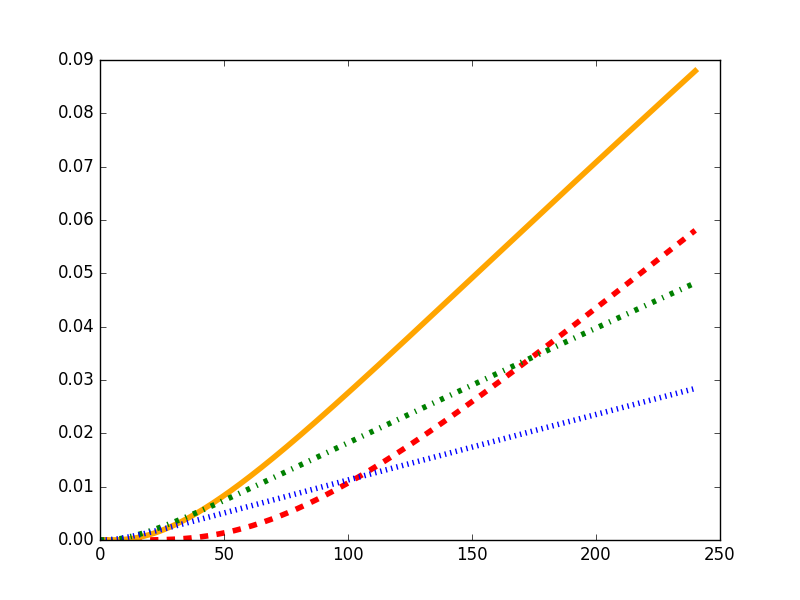}
\put(34,-2){Time (min) }
\put(-6,20){\rotatebox{90}{$P(x_B,y_B,t)$}}
\end{overpic} } 
\caption{Probability of infection versus time in an 8m $\times$ 8m $\times$ 3m classroom for an infectious person breathing in a case of very poor ventilation (red dashed), poor ventilation (orange solid), pre-pandemic recommended ventilation (green dot-dashed) and pandemic-updated recommended ventilation (blue dotted), using \eqref{eq:indoors-sol} and \eqref{eq:prob}, with parameter values as given in Tables \ref{tab:parameters} and \ref{tab:local-parameters}. (a) Evaluated at Position $A$:(5,4). (b) Evaluated at Position $B$:(8,8). Here, $\mathcal{R}=1$, which corresponds to  breathing.}
\label{fig:Prob-vs-t}
\end{figure}

Figure~\ref{fig:Prob-vs-t} shows the probability of infection versus time evaluated at Position $A$ and Position $B$ in the case study 8m $\times$ 8m $\times$ 3m classroom for the case of an infectious person just breathing. In Figure~\ref{fig:Prob-vs-t-XY88}, we can see the effect of the concentration building more slowly at very poor ventilation that we saw in Figure~\ref{fig:C-vs-t-XY88}, with the probability of infection growing very slowly initially then overtaking the two better ventilation scenarios 3 and 4. As Position $A$ and Position $B$ are the downwind locations in the room with the highest and lowest concentration, they are also the positions with the highest and lowest infection risk in the room. Hence, by examining these two points we have a range of the risk in the room.

\subsubsection{Average Probability of Infection}
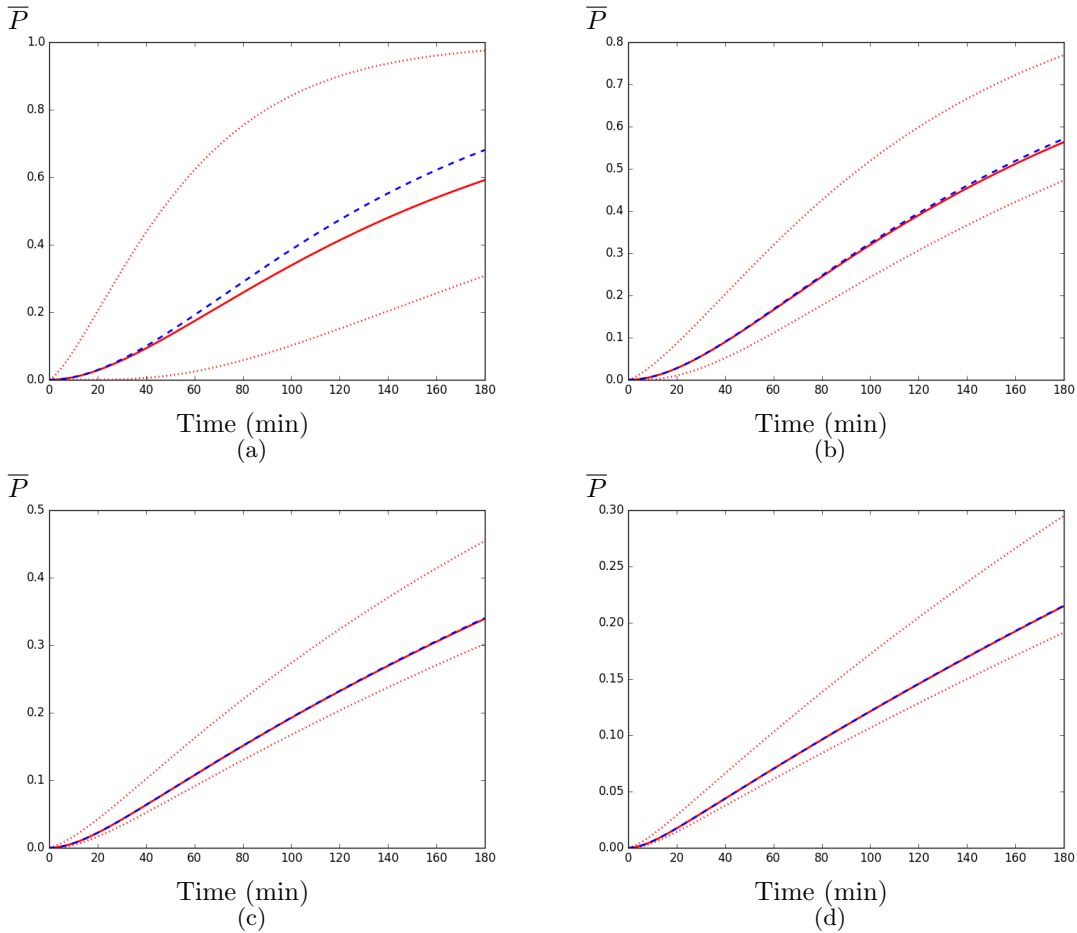
\begin{figure}[h]
\centering
\subfloat[\label{fig:avg-compare-Q0}]{
\begin{overpic}[width=.45\textwidth,tics=10]{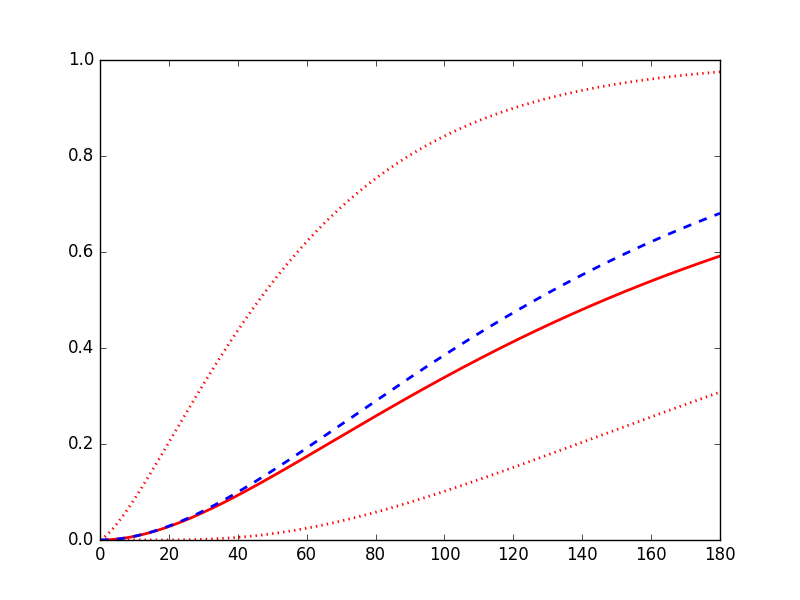}
\put(35,-2){Time (min) }
\put(5,70){$\overline{P}$}
\end{overpic} }
\subfloat[\label{fig:avg-compare-Q1}]{
\begin{overpic}[width=.45\textwidth,tics=10]{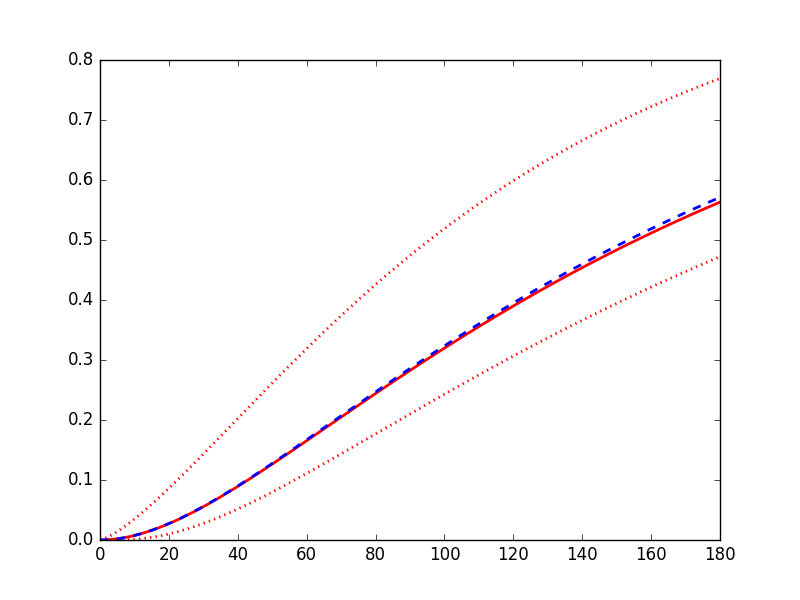}
\put(35,-2){Time (min)}
\put(5,70){$\overline{P}$}
\end{overpic} } \\
\subfloat[\label{fig:avg-compare-Q2}]{
\begin{overpic}[width=.45\textwidth,tics=10]{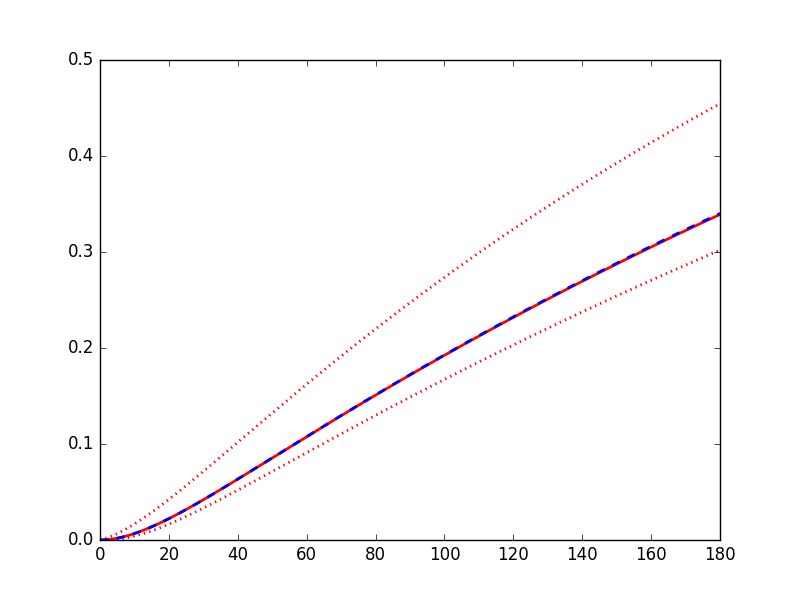}
\put(35,-2){Time (min) }
\put(5,70){$\overline{P}$}
\end{overpic} }
\subfloat[\label{fig:avg-compare-Q3}]{
\begin{overpic}[width=.45\textwidth,tics=10]{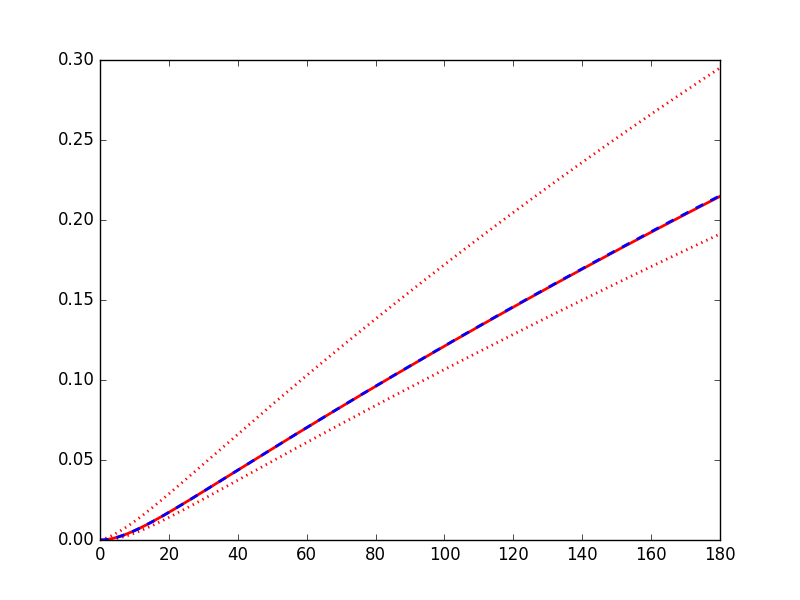}
\put(35,-2){Time (min) }
\put(5,70){$\overline{P}$}
\end{overpic} }
\caption{Comparison of spatially averaged probability of infection $\overline{P}$ (red solid) versus probability of infection from the spatially averaged concentration (blue dashed) in a 8m $\times$ 8m $\times$ 3m classroom with an infectious person talking, from \eqref{eq:indoors-sol} and \eqref{eq:prob}, with parameter values from Tables \ref{tab:parameters} and \ref{tab:local-parameters}: (a) very poor ventilation, (b) poor ventilation (c) pre-pandemic recommended ventilation (d) pandemic-updated recommended ventilation. Here,  $\mathcal{R}=10$, which corresponds to talking.}
\label{fig:avg-compare}
\end{figure}

In Figure~\ref{fig:avg-compare}, we present a comparison of the spatially averaged probability of infection versus the probability of infection from the average concentration in our model for the case of a person talking in the 8m $\times$ 8m $\times$ 3m classroom. As Figures \ref{fig:avg-compare}b--d show, as ventilation improves, the average probability is almost equal to the probability derived from the average concentration. However, for the very poor ventilation scenario, Figure~\ref{fig:avg-compare-Q0} shows that the spatially averaged probability is lower than the probability from the average concentration in the room. This figure demonstrates that our spatially dependent model more accurately captures the infection risk than the spatially averaged (WMR) approximations. 

\newpage
\subsection{Case Study: Restaurant (Comparison with CFD models)}
\label{Case Study: Restaurant}

We compare our model to the CFD models in \cite{Li2020}, \cite{Birnir2020} and \cite{Ho2021rest}. These models studied a well-known superspreader outbreak in a restaurant in Guangzhou, China that occurred on January 23, 2020. The restaurant had no outdoor air supply and all the ventilation was supplied by five air conditioning units, four of which were mounted on the same wall, while the fifth was at the opposite wall from the outbreak region in the restaurant--see Figure 1 in \cite{Li2020}. 7--9 people out of 20 seated in that part of the restaurant were infected. This has been associated with a recirculating airflow from one of the AC untis \cite{Lu2020}. We assumed that the recirculation of air in  the restaurant due to the AC units is similar to the air recirculation we assumed in our model (see Figure~\ref{fig:recirculation}) and generate predictions on the particle concentration and the infection risk which sufficiently agree with the CFD models.

The restaurant has an almost trapezoidal shape, with the length of the opposite walls 6 and 8.3\,m, respectively and height 3.14\,m (hence, volume of 431\,m$^3$)--see Figure 1 in \cite{Li2020}. We decide to simulate the restaurant using a rectangular domain of $l=6$\,m, $w=17$\,m and $h=3.14$\,m, because the length of the part of the restaurant where the outbreak occurred was 6\,m. Tracer gas experiments from \cite{Li2020} showed that the air exchange rate, $\lambda$ was $0.77\,\rm{h}^{-1}$, which corresponds to poor ventilation. 

\begin{figure}[h]
\centering
\subfloat[\label{fig:C-rest-t1}]{
\begin{overpic}[width=.5\textwidth,tics=10]{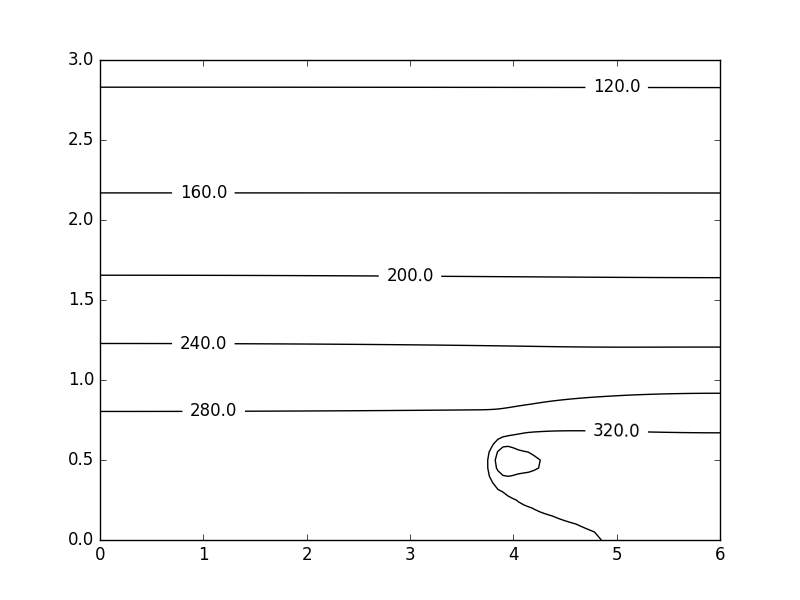}
\put(43,-1){$x$ (m) }
\put(6,71){$y$ (m)}
\end{overpic} }
\subfloat[\label{fig:C-rest-avg}]{\begin{overpic}[width=0.5\textwidth, tics=10]{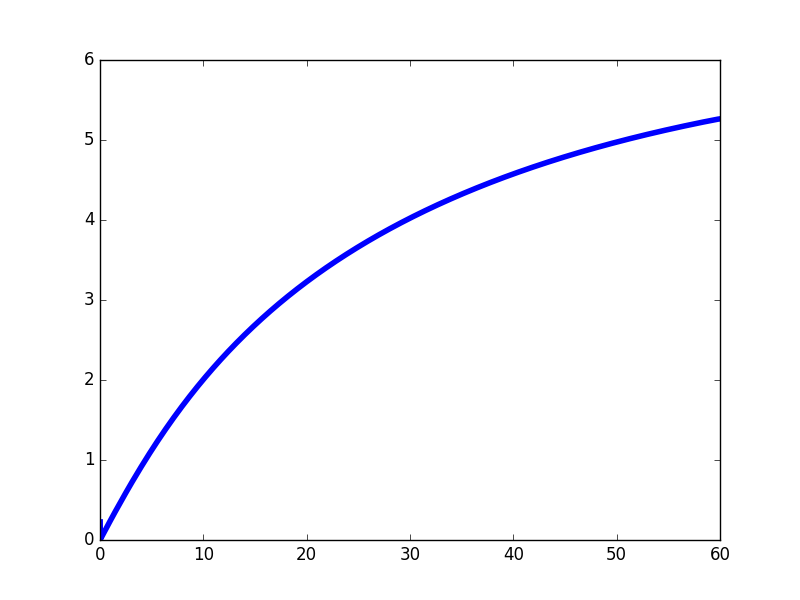}
    \put(37,-1){Time (min)}
    \put(9,70){$\overline{C}$}
    \end{overpic}
    }
\caption{(a) Concentration of SARS-CoV-2-carrying particles after 1 hour, (b) Spatially averaged concentration $\overline{C}$ normalised by the concentration 0.5\,m downwind from the infectious person when there is no recirculation for the outbreak region in the Guangzhou restaurant superspreader outbreak in \cite{Lu2020}, obtained from \eqref{eq:indoors-sol} with $l=6$m, $w=17$m, $h=3.14$m, $\lambda = 0.77\rm{h}^{-1}$, $V=431$\,m$^3$, $N=5$, $(x_0,y_0)=(3.75,0.5)$. Other parameter values as given in Table \ref{tab:parameters}. Here, $\mathcal{R} = 10$, which corresponds to talking.}
\label{fig:CFD-compare}
\end{figure}

To obtain the value of $K$, we use \eqref{eq:K} with $V=431$\,m$^3$, $\lambda = 0.77$\,h$^{-1}$, and the number of air conditioning units, $N=5$. As this was a social gathering in a restaurant, we also assume that $\mathcal{R}=10$ for talking and that the susceptible people had a sedentary breathing rate $\rho = 1.3\times10^{-4}$\,m$^3$/s.
The specific location of the infectious person based on the figure provided in \cite{Lu2020} is taken to be  $(x_0,y_0)$ is $(3.75,0.5)$. (This location was not provided by \cite{Li2020} or \cite{Lu2020}.)
As the infectious person was in the restaurant for 1 hour and 20 minutes, and infected diners shared the room for 50--75 minutes \cite{Li2020}, we run the simulation for one hour. The rest of the parameters used can be found in Table~\ref{tab:parameters}. We present the concentration in the room after 1 hour predicted by our model, showing only the outbreak region of the restaurant  in Figure~\ref{fig:CFD-compare}.

In \cite{Li2020}, they found the air flows from the five air-conditioning units in the restaurant formed `recirculation zones'. In the recirculation zone where the infectious person was, the air-conditioning stream carried the infectious aerosols to the opposite window, then the stream bent downward and returned at a lower height before finally the contaminated air rose and returned to the air-conditioning unit, repeating the cycle \cite{Li2020}. The dose of infectious particles in this region normalised by the dose at the table of the infectious person obtained by \cite{Li2020} and our model is presented in Table~\ref{tab:CFD-compare}.
To make this comparison in Table~\ref{tab:CFD-compare}, we again assumed from the Figure in \cite{Lu2020} that the downwind table coordinates are $(5,1)$, the upwind table coordinates are $(1,1)$, and the adjacent table outside the recirculation zone is at $(3.5,3)$. We normalised these results by the dose at $(3.5,0.5)$, which we chose to approximate the coordinates of the table where the infectious person is sitting at.

\begin{table}[ht]
    \centering
    \caption{Normalised concentration in the restaurant after an hour from \cite{Li2020}, adapted from Table 3 from \cite{Li2020}, and our model, from \eqref{eq:indoors-sol} and \eqref{eq:dose} with $l=6$m, $w=17$m, $h=3.14$m, $\lambda = 0.77\rm{h}^{-1}$, $V=431$\,m$^3$, $N=5$, $(x_0,y_0)=(3.75,0.5)$ and other parameter values as given in Table \ref{tab:parameters}. Here, $\mathcal{R} = 10$, which corresponds to talking.}
    \begin{tabular}{|c|c|c|} \hline
        Table & Normalised dose in \cite{Li2020} & Our Normalised dose \\ \hline
        Downwind table & 0.76 & 0.94 \\ \hline
        Upwind table & 0.89 & 0.89 \\ \hline
        Adjacent table & 0.40 & 0.34  \\ \hline
    \end{tabular}
    
    \label{tab:CFD-compare}
\end{table}

Let's discuss further the simplifying assumption that our recirculation flow is similar to that in the recirculation zones in \cite{Li2020}. The result in Table~\ref{tab:CFD-compare} shows the normalised dose predicted by our model at the adjacent table outside the recirculation zone is slightly less than that predicted by \cite{Li2020}, so the effect of this recirculation zone is still present in our model. This occurs because the diffusion rate of the infectious particles in our model in this scenario is low ($3.3\times 10^{-3}\,\rm{m}^2/\rm{s}$). The results in Table~\ref{tab:CFD-compare} shows good agreement between the models for the upwind table. However, \cite{Li2020} predicts a lower concentration downwind than upwind, which our model cannot do.

Similarly, \cite{Birnir2020} and \cite{Ho2021rest} showed that their simulated recirculation of infectious aerosols was mainly confined to this region of the room, hence the concentration was significantly higher in this region compared to the rest of the room. The CFD simulations in \cite{Birnir2020} showed that in the region of the outbreak the average concentration after 12 minutes is as high as the concentration at the seat directly next to the infected person and after 60 minutes is 5 times higher than the concentration next to the person, if there was no recirculation. In Figure~\ref{fig:C-rest-avg}, we present the average concentration predicted by our model in this region normalised by the concentration 0.5\,m downwind from the infectious person when there is no recirculation, which we chose to simulate a person sitting directly next to the infectious person. Our results show that after 60 minutes, the average concentration is 5 times higher. Therefore, our model shows good agreement with the results of \cite{Birnir2020}.

\begin{figure}[h]
    \centering
\subfloat[\label{fig:prob-rest-contour}]{
\begin{overpic}[width=.48\textwidth,tics=10]{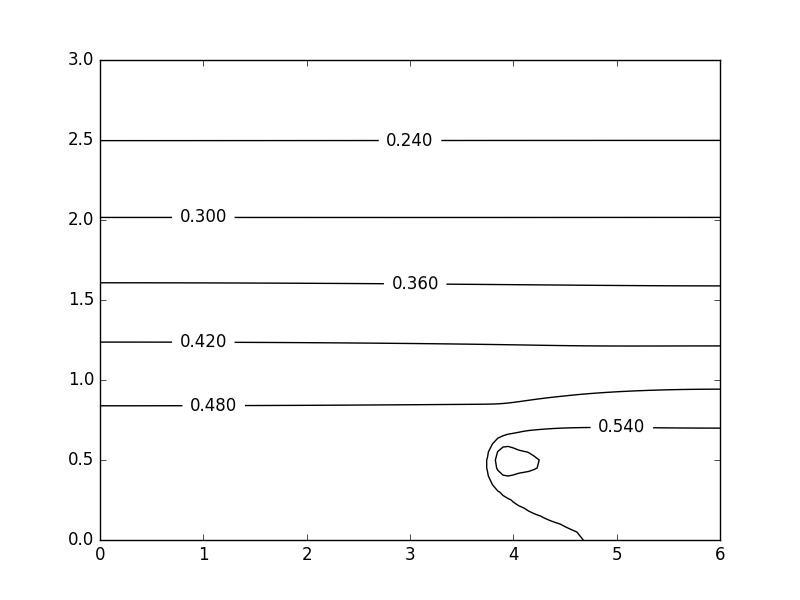}
\put(43,-2){$x$ (m) }
\put(4,73){$y$ (m)}
\end{overpic} }
\subfloat[\label{fig:prob-rest-average}]{
\begin{overpic}[width=.48\textwidth,tics=10]{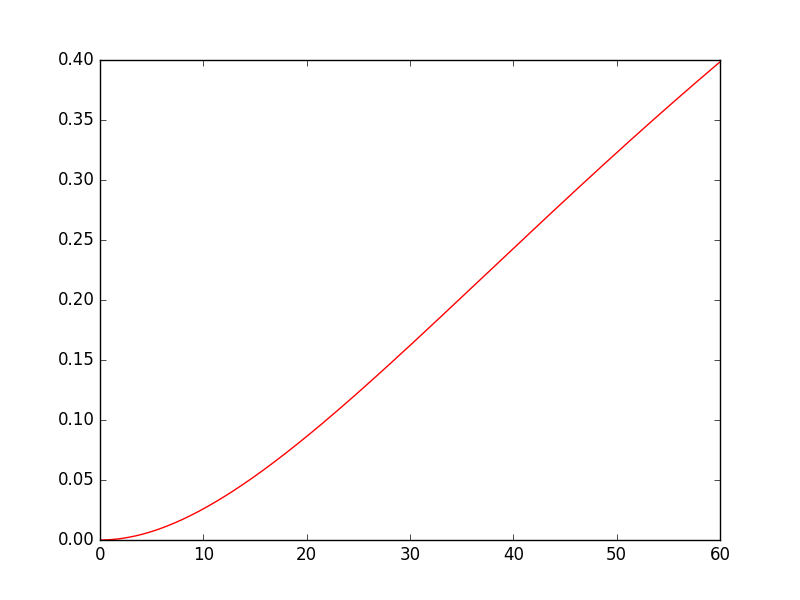}
\put(30,-2){Time (minutes)}
\put(8,73){$\overline{P}$}
\end{overpic} }
    \caption{(a) Probability of infection after one hour and (b) Average probability of infection $\overline{P}$ versus time for the outbreak region in the Guangzhou restaurant superspreader outbreak described in \cite{Lu2020}, from \eqref{eq:indoors-sol} and \eqref{eq:prob}, $l=6$m, $w=17$m, $h=3.14$m, $\lambda = 0.77\rm{h}^{-1}$, $N=5$, $(x_0,y_0)=(3.75,0.5)$ and other parameter values as in Table~\ref{tab:parameters}. Here, $\mathcal{R} = 10$, which corresponds to talking.}
    \label{fig:prob-restaurant}
\end{figure}

 In this outbreak, 7--9 out of 20 people in this part of the restaurant were infected \cite{Lu2020}, which corresponds to an infection rate of 35-45\%. The CFD simulation from \cite{Ho2021rest} produced a probability of infection of between 15-40\% in this region, using a low viral load. The probability of infection for this restaurant, generated by our model, is shown in Figure~\ref{fig:prob-restaurant}. From Figure~\ref{fig:prob-rest-contour}, we infer that this region of the restaurant experienced a probability of infection of 25--65\%, which leads to an average probability of 40\% (see Figure~\ref{fig:prob-rest-average}). 
 Therefore, the probability results from our model are in good agreement with this outbreak.

\subsection{Time to Probable Infection (TTPI)}
\label{sec:TTPI}
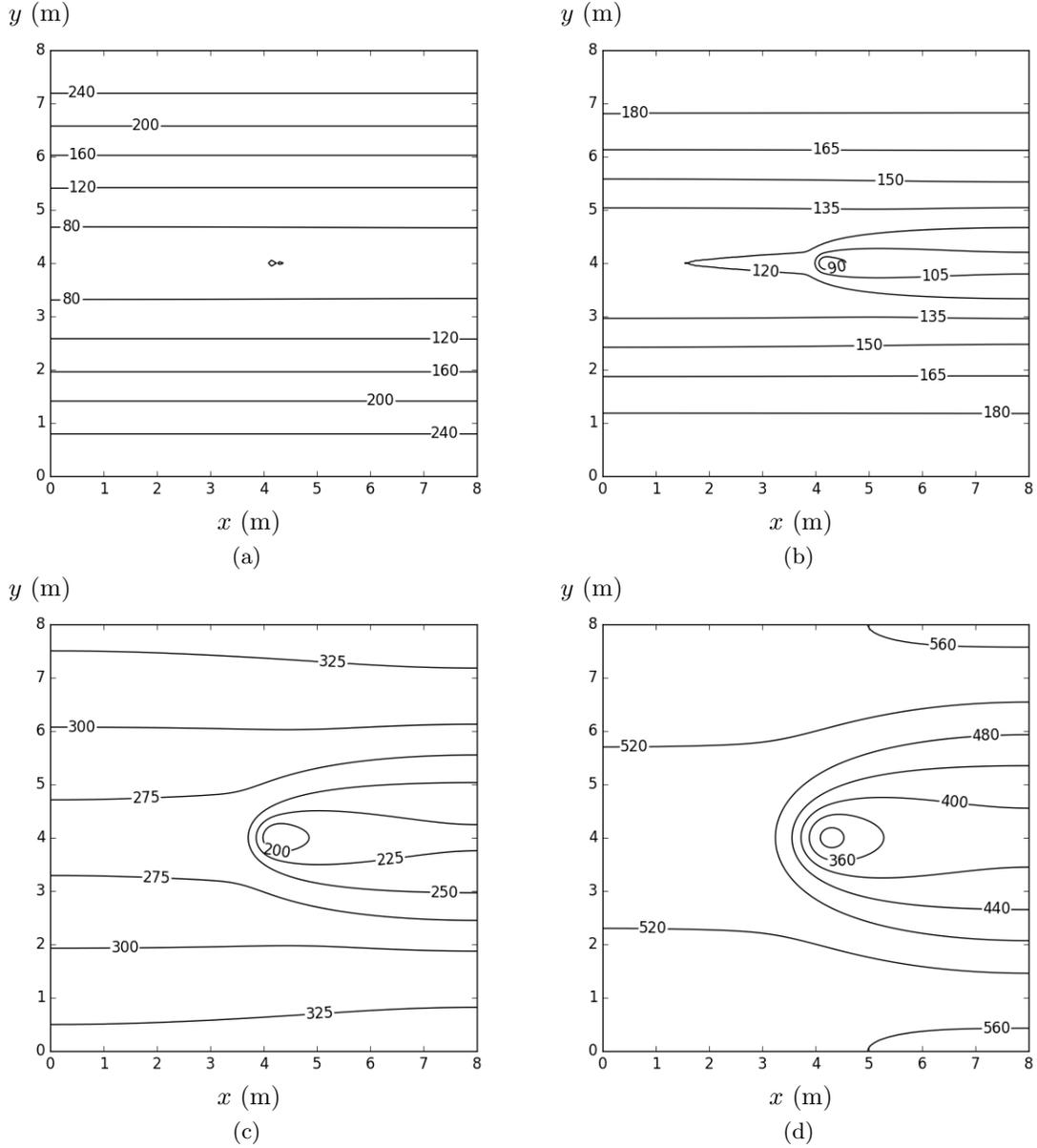
\begin{figure}[p]
\centering
\subfloat[\label{fig:TTPI-Q0}]{
\begin{overpic}[trim = 2.5cm 0 2.5cm 0, clip,width=.45\textwidth,tics=10]{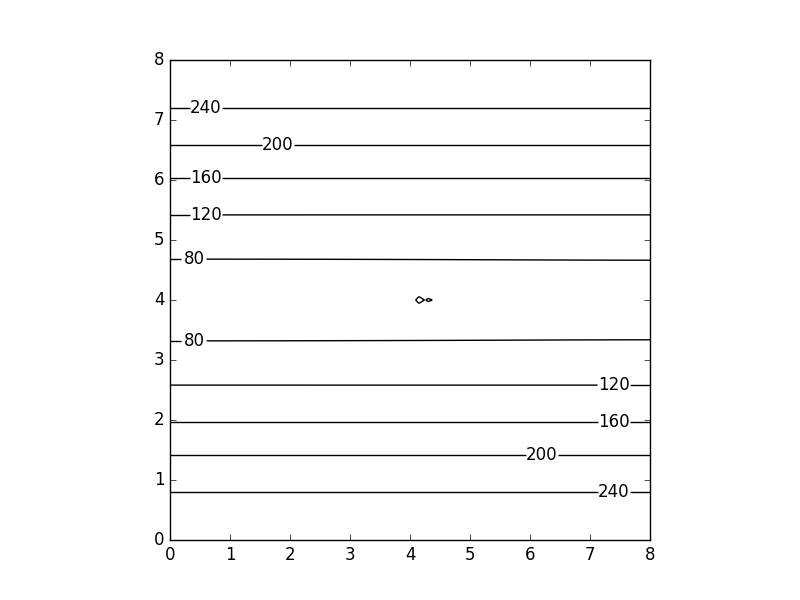}
\put(43,0){$x$ (m) }
\put(4,95){$y$ (m)}
\end{overpic} }
\subfloat[\label{fig:TTPI-Q1}]{
\begin{overpic}[trim = 2.5cm 0 2.5cm 0, clip,width=.45\textwidth,tics=10]{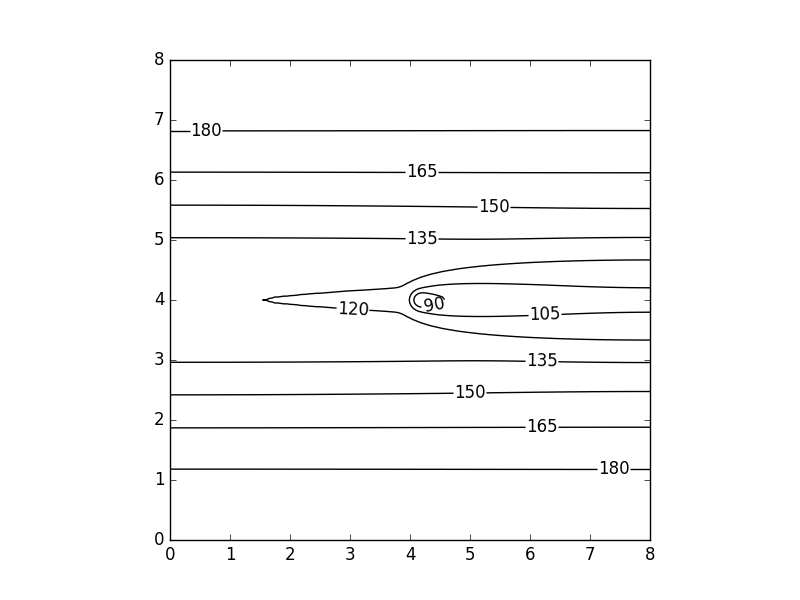}
\put(43,0){$x$ (m) }
\put(4,95){$y$ (m)}
\end{overpic} } \\
\subfloat[\label{fig:TTPI-Q2}]{
\begin{overpic}[trim = 2.5cm 0 2.5cm 0, clip,width=.45\textwidth,tics=10]{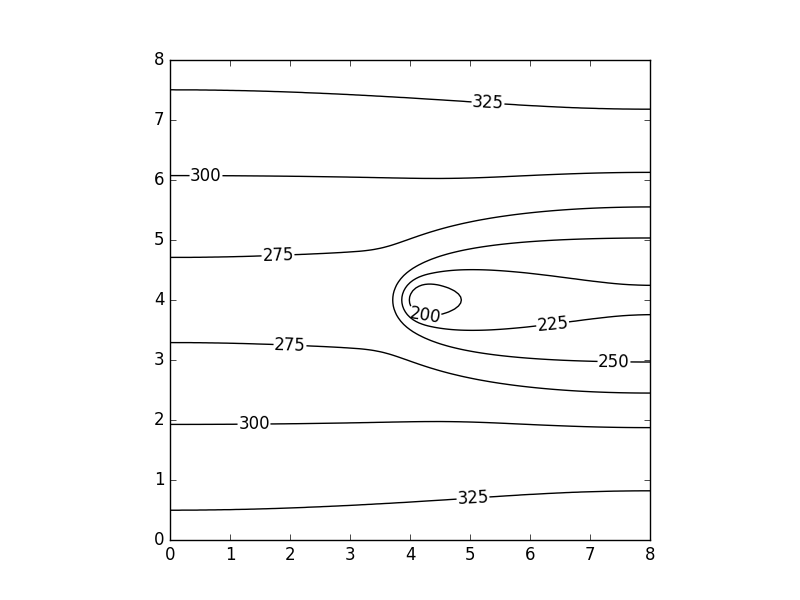}
\put(43,0){$x$ (m) }
\put(4,95){$y$ (m)}
\end{overpic} }
\subfloat[\label{fig:TTPI-Q3}]{
\begin{overpic}[trim = 2.5cm 0 2.5cm 0, clip,width=.45\textwidth,tics=10]{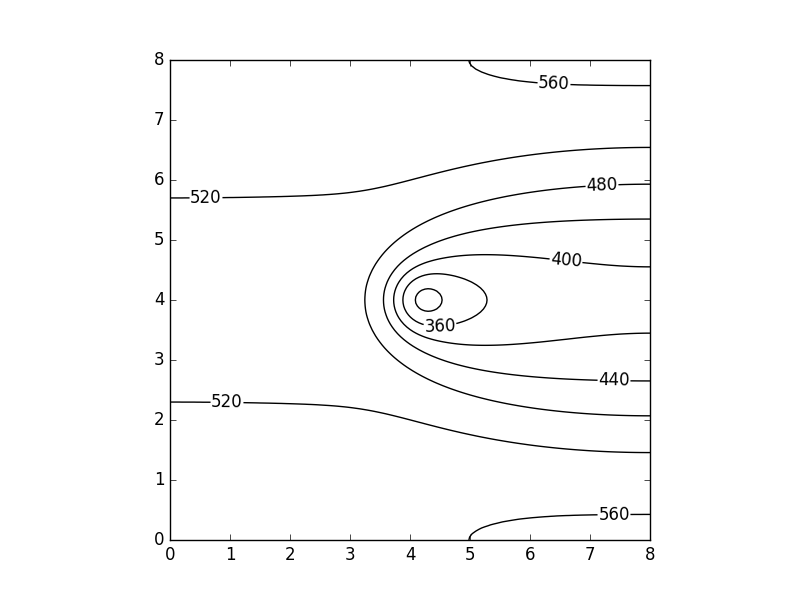}
\put(43,0){$x$ (m) }
\put(4,95){$y$ (m)}
\end{overpic} }
\caption{`Time to probable infection’  (TTPI) due to an infectious person talking at the centre of a 8m $\times$ 8m $\times$ 3m room from \eqref{eq:indoors-sol} and \eqref{eq:prob}, with parameter values from \ref{tab:parameters} and \ref{tab:local-parameters}: (a) very poor ventilation, (b) poor ventilation (c) pre-pandemic recommended ventilation (d) pandemic-updated recommended ventilation. $\mathcal{R} = 10$ for talking.}
\label{fig:TTPI-contour}
\end{figure}

Paving the way towards formulating policy recommendations, we can also use the model to calculate the `time to probable infection’ (TTPI) -- the time required for the infection risk at a point in the room to reach 50\%. We present in Figure~\ref{fig:TTPI-contour} the spatially dependent TTPI for the case of an infectious person talking in the 8m $\times$ 8m $\times$ 3m classroom, previously considered.

\begin{figure}[t]
\centering
\subfloat[\label{fig:SOT-vs-Q-XY54}]{
\begin{overpic}[width=.5\textwidth,tics=10]{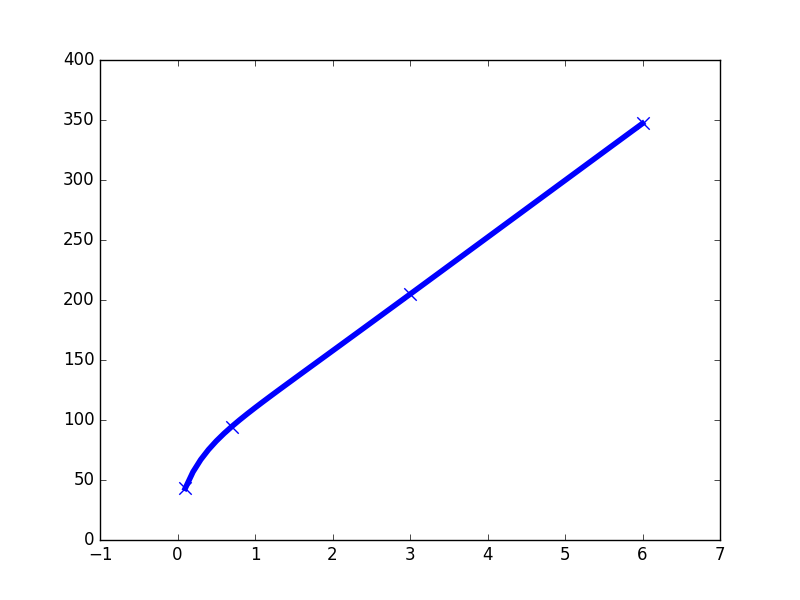}
\put(45,-2){$\lambda$ ($\rm{h}^{-1}$) }
\put(-10,70){$TTPI$ (minutes)}
\end{overpic}}
\subfloat[\label{fig:SOT-vs-Q-XY88}]{
\begin{overpic}[width=.5\textwidth,tics=10]{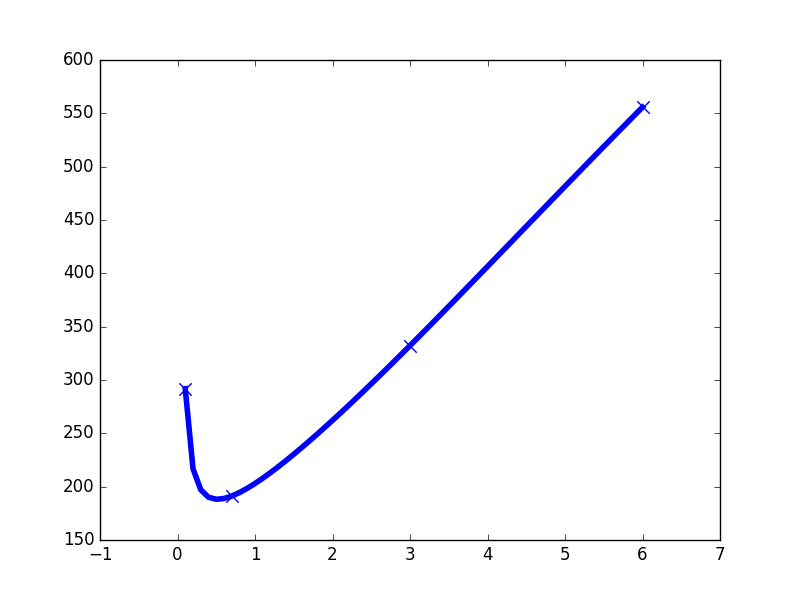}
\put(45,-2){$\lambda$ ($\rm{h}^{-1}$) }
\put(-10,70){$TTPI$ (minutes)}
\end{overpic} }
\caption{The TTPI versus the air exchange rate $\lambda$ for an infectious person talking in the room; from \eqref{eq:indoors-sol} and \eqref{eq:prob}: (a) at Position $A$:(5,4) (b) at Position $B$:(8,8). All parameter values are given in Tables \ref{tab:parameters} and \ref{tab:local-parameters}. Here, $\mathcal{R} = 10$, which corresponds to talking.}
\label{fig:SOT-vs-Q}
\end{figure}

Figure \ref{fig:SOT-vs-Q} shows the relationship between the TTPI and $\lambda$ at Positions $A$ and $B$ in our classroom case study, respectively. TTPI monotonically increases with $\lambda$. We can extract the following values for the four ventilation scenarios we consider. For very poor ventilation TTPI$\approx 50$ minutes and for poor ventilation TTPI$\approx100$ minutes. For the latter, better ventilation scenarios, TTPI exceeds $200$ and $350$ minutes, respectively. In Figure \ref{fig:SOT-vs-Q-XY54} we observe further evidence of the scaling-law behaviour at Position $A$ that was identified in Figure~\ref{fig:C-vs-t-XY54}, by now finding that TTPI satisfies the relationship TTPI\,$\propto \lambda^\gamma$, where $\gamma \approx 0.41$ for $\lambda < 1 $h$^{-1}$, and a linear relationship, i.e. $\gamma \approx 1$, for $\lambda > 1 $h$^{-1}$. We note that as Point $A$ is the most dangerous point in the room with the lowest TTPI (while maintaining 1m social distancing), the TTPI at $A$ could be used to determine a `safe occupancy time' for the whole room. (Recall that we do not generally know where the infectious person is located in the room.)

At Position $B$, Figure \ref{fig:SOT-vs-Q-XY88}, at the top right corner we see that  the TTPI depends non-monotonically on $\lambda$ with the TTPI attaining a minimum value (most dangerous point) for $\lambda$ approximately equal to 0.6$h^{-1}$. For values of $\lambda$ smaller than 0.6$h^{-1}$ the TTPI \emph{decreases} with $\lambda$; for values of $\lambda>0.6 h^{-1}$ TTPI again increases with $\lambda$. At first glance, this is perhaps counter-intuitive since it implies that more ventilation makes the room less safe but it can be explained by the slower initial build-up of concentration for this range of $\lambda$ (see Figure \ref{fig:C-vs-t-XY88}. Crucially, this non-monotonic behaviour of TTPI demonstrates that very low ventilation could increase the infection risk in some parts of the room away from the source so a very important take-away from our work is that `increasing the ventilation a little could actually be worse'. This supports the initial observation made in Figure~\ref{fig:C-vs-t}. 

We note that Figures \ref{fig:TTPI-contour} and \ref{fig:SOT-vs-Q} could be generated very quickly for any location, parameters and ventilation settings of interest.

\begin{figure}[hb]
\centering
\subfloat[\label{fig:SOT-vs-R-XY54}]{
\begin{overpic}[width=.5\textwidth,tics=10]{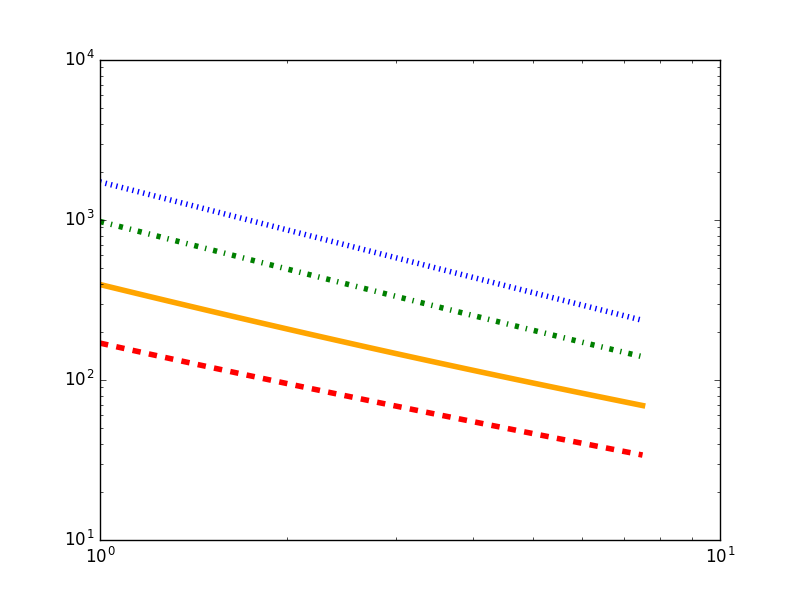}
\put(40,-1){$\log(R)$}
\put(2,71){$\log$(TTPI)}
\end{overpic} }
\subfloat[\label{fig:SOT-vs-R-XY88}]{
\begin{overpic}[width=.5\textwidth,tics=10]{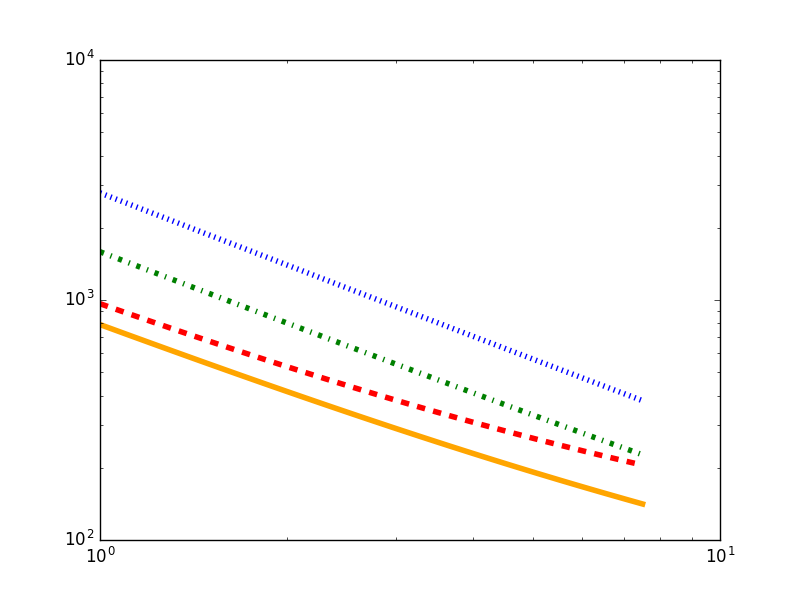}
\put(40,-1){log(R)}
\put(2,71){log(TTPI)}
\end{overpic} }
\caption{Log--log plot of the TTPI versus $R$ in a case of very poor ventilation (red dashed), poor ventilation (orange solid), pre-pandemic recommended ventilation (green dot-dashed) and pandemic-updated recommended ventilation (blue dotted): (a) at Position $A$, (b) at Position $B$, obtained from \eqref{eq:indoors-sol} and \eqref{eq:prob}. All parameter values are given in Tables \ref{tab:parameters} and \ref{tab:local-parameters}.}
\label{fig:SOT-vs-R}
\end{figure}

\newpage
Next, in order to quantify the dependence of TTPI on the type of activity we plot in Figure \ref{fig:SOT-vs-R} the TTPI vs $R$, the emission rate of airborne infectious particles, for the four ventilation scenarios. At Position $A$ there is a power law of the form TTPI\,$\propto R^{\mu}$ with $\mu\approx$ -0.79, -0.85, -0.96 and -0.99, respectively. At Position $B$, $\mu\approx$ -0.75, -0.84, -0.96 and -0.99, respectively. As expected, the TTPI decreases as $R$ increases, at both $A$ and $B$. As $\mu$ decreases with increasing ventilation, the dependence of TTPI on $R$ becomes weaker with increasing ventilation. In other words, the more particles generated per second the less safe the room is but when ventilation increases sufficiently the type of activity matters less. We also note that at Position $B$, for the poorer ventilation settings, the effects of the weaker mixing are visible again. Furthermore, as $\rho$ and $I$ have the same influence in \eqref{eq:prob} as $R$, we could also determine analogous power laws of the forms TTPI\,$\propto \rho^{\epsilon}$ or TTPI\,$\propto I^{\nu}$.

\section{Conclusions and Future Directions}
\label{Section:Summary and Conclusions}
In this paper we have addressed the need to develop a quick-to-run and efficient model that determines the spatially dependent concentration of airborne particles (aerosols) carrying the SARS-CoV-2 virus and also the corresponding probability of infection (infection risk). The model can be applied to a variety of indoor locations.

We placed one infectious, asymptomatic or presymptomatic person in the centre of the room who either breathes or talks, with or without a mask; this corresponds to four different emission rates for the viral particles (Table 1). The semi-analytic solution allows for fast simulations; for example, it takes only one minute of simulation to determine the concentration after one hour of real time, on a standard laptop. Subsequently, inserting the spatially varying concentration \eqref{eq:indoors-sol} in the well-known exponential probability density function of the Wells--Riley models \cite{Riley1978,Watanabe2010,burridge2021} we determined, for the first time, a spatially varying probability of infection \eqref{eq:prob}. 

The model relies on many parameters that are currently unknown for SARS-CoV-2 or which are known to vary over a wide range. Particularly uncertain parameters are the infectious dose, the infectiousness constant, the viral load and the length of time before the virus becomes biologically inactivated. We have used the best estimates for these parameters by consulting the literature and talking to scientists involved in experiments. However, our predictions must still be taken with caution. We do believe, however, that our model provides an important contribution to the COVID-19 modelling field since it can be easily updated and simulated if and when more accurate estimates are obtained. The model can also be used to quickly determine the comparative infection risk arising from new variants of the virus (see \eqref{eq:prob-scale}) or from different types of activity, such as physical exercise (see \eqref{eq:prob-scale-rho}).

In Section \ref{Case Study: A classroom} we applied the model to an average-sized classroom, with dimensions 8m $\times$ 8m $\times$ 3m (Figure 3), for four different ventilation settings, characterised by $\lambda$, the air exchange rate. Our model shows that the concentration, and the infection risk in the room are highest downwind from the infectious person (see Figure~\ref{fig:C-contour}). As in the Wells--Riley model, the concentration in the room increases initially, then reaches a steady state (see Figure~\ref{fig:C-vs-t}). We consider the concentration at two indicative points: A (1m downstream of the source, assuming social distancing); and B (top right corner). At point A the time to reach the steady state is shorter and the steady-state concentration is lower for larger values of $\lambda$. However, at point B the highest concentration is attained in the poor ventilation scenario. Moreover, the better ventilation scenarios 3 and 4 exhibit higher concentrations than the very poor ventilation for some time before the poor ventilation surpasses these values. At position A the concentration obeys a power-law dependence on time to begin with, since the spreading behaviour is unimpeded by obstacles. However, as time progresses and the effect of the walls comes into play, the power law breaks down. Analogously, no power law exists for B, since the influence of the walls is significant for all time due to the proximity of the walls to this location.

Another observation from Figure~\ref{fig:C-contour} is that the second highest concentration level is found upwind of the source (on the same horizontal line). Our results agree qualitatively with air sampling data from hospital wards in Wuhan \cite{Guo2020}, which showed that virus-carrying particles were `mainly concentrated near and downstream from the patients' and there was also an `exposure risk upstream'.

Our model can also be used to find the \emph{vacancy time} required to reset the viral concentration in classrooms or other locations where people stay for a prolonged amount of time, for example offices. In Figure~\ref{fig:C-vs-t-lunch}, we present the concentration at Position $A$ and Position $B$ in the room for a seven-hour day with a one-hour lunch break in the middle of the day. For the ASHRAE-recommended scenarios 3 and 4, a one-hour vacancy is shown to be sufficient to reset the concentration. However, for the poor and very poor ventilation scenarios 1 and 2, an hour is insufficient to clear the room.

In Figure \ref{fig:Prob-contour} we show the probability of infection for the four ventilation settings we consider (8m $\times$ 8m $\times$ 3m classroom, infectious person breathing for one hour). The contours in Figure~\ref{fig:Prob-contour} are similar in shape to the concentration contours in Figure~\ref{fig:C-contour}. This implies that the greatest risk of infection indoors is directly downwind from the infectious person, and the risk decreases as we travel away from the source in a direction orthogonal to the airflow. Figure~\ref{fig:Prob-vs-t} shows the probability of infection versus time evaluated at Positions $A$ and $B$. In Figure~\ref{fig:Prob-vs-t-XY88}, we can see the effect of the concentration building more slowly for very poor ventilation (Figure~\ref{fig:C-vs-t-XY88}), with the probability of infection growing very slowly initially then surpassing the two better ventilation scenarios 3 and 4. As Position $A$ and Position $B$ are the downwind locations in the room with the highest and lowest concentrations, respectively, they are also, respectively, the downwind positions with the highest and lowest infection risk in the room. Hence, by examining these two points, we have a range of the downstream risk in the room. 

In Figure~\ref{fig:avg-compare}, we present a comparison of the spatially averaged probability of infection versus the probability of infection obtained from the average concentration arising from our model. The latter infection risk is that predicted by previous, spatially uniform Wells--Riley models. We consider an infectious person talking, in the 8m $\times$ 8m $\times$ 3m classroom. As Figures \ref{fig:avg-compare}b--d show, as ventilation improves, the average probability becomes almost equal to the probability derived from the average concentration. However, for the very poor ventilation scenario, Figure~\ref{fig:avg-compare-Q0} shows that the spatially averaged probability is lower than the probability from the average concentration in the room. This figure suggests that the spatially averaged (WMR) approximations may overestimate the infection risk for poorer ventilation scenarios.

We also compared our model to the CFD models in \cite{Li2020}, \cite{Birnir2020} and \cite{Ho2021rest} simulating a superspreader outbreak in a restaurant in Guangzhou, China (January 23, 2020), with good agreement. The outbreak infected 7--9 people \cite{Lu2020} out of the 20 in the particular region of the restaurant. Hence, the average infection risk of 40$\%$ we predict with the model is a satisfactory prediction.

Paving the way for formulating policy-making recommendations, in Section \ref{sec:TTPI} we used the spatially varying concentration to determine the `time to probable infection’ (TTPI) as a function of $\lambda$. In the downwind region, 1m away from the source (point A), the TTPI increases with $\lambda$, exhibiting a scaling-law dependence on $\lambda$ (see Figure~\ref{fig:SOT-vs-Q-XY54}). Outside this high-concentration region (point B), the TTPI, interestingly, exhibits a non-monotonic relationship with $\lambda$, first decreasing to a minimum TTPI and then increasing again with $\lambda$ (see Figure~\ref{fig:SOT-vs-Q-XY88}). This observation ties up with previous observations that improving ventilation a little may be worse than not doing anything.

We have also identified a power-law relation between the TTPI and $R$, the emission rate of airborne infectious particles (see Figure~\ref{fig:SOT-vs-R}) that encapsulates the TTPI over a continuous range of $R$. Consequently, this can be used to quickly quantify the effect on the TTPI of the activity type for one person (for example talking at low or high volume) or differences in the emission rate between persons (as some people emit more viral particles than others for the same type of activity). Analogous power laws can be determined for the TTPI versus $I$, the infectiousness constant in the probability density function \eqref{eq:prob}; and for the TTPI versus $\rho$, the breathing rate of the susceptible person. These would allow us to quickly generate predictions for new variants, once the parameters are made available.

The model could be extended in several ways to account for additional questions of interest. Currently we are working on including viral particle size distributions, particularly size-dependent gravitational settling and emission rates that depend on the particle size. We are also extending this model for rooms with more complex geometries, air velocities and ventilation systems. Of particular interest is to quantify the effect to the infection risk when air purifiers are introduced in rooms, a stream of work motivated by our industrial partner Smart Separations Ltd.

In conclusion, our model can be implemented to assess the risk of airborne infection in rooms of different sizes when an infectious person is breathing or talking with or without a mask. It can also be implemented to quantify the effectiveness of changing the ventilation available in the room. The modelling framework contains a series of parameters that may easily be adjusted in light of new guidelines on ventilation or confirmed information on the virus.  Most importantly for the current COVID-19 pandemic, it can be implemented quickly on individual, widely available and inexpensive computers. 

\section*{Acknowledgements}
KK and AE gratefully acknowledge funding from a S$\hat{\textrm{e}}$r Cymru COVID-19 grant, awarded by the Welsh government. IMG gratefully acknowledges support from the Royal Society via a University Research Fellowship. We are grateful to Si$\hat{\textrm{a}}$n Grant who generated Figures 1 and 2a. We are also grateful to Raquel Gonz\'{a}lez Fari\~na and Alexander Ramage (Cardiff University) who enthusiastically joined the team and have been participating in many fruitful discussions. We thank Hugo Macedo, Carina Dos Santos and Sotiria Tsochataridou from Smart Separations Ltd for valuable discussions.

\bibliographystyle{unsrt}
\bibliography{references}

\end{document}